\documentclass[]{article}

\usepackage[in]{fullpage}	
\usepackage{graphicx, amsmath, amssymb, xfrac}
\usepackage{url}
\usepackage[letterpaper=true,colorlinks=false]{hyperref}
\usepackage{bookmark}	
\usepackage{multirow} 

\def\nn{\nonumber}

\newcommand{\pd}[2]{\frac{\partial{#1}}{\partial{#2}}}
\newcommand{\spd}[2]{\frac{\partial^2{#1}}{\partial{#2}^2}}
\newcommand{\td}[2]{\frac{d{#1}}{d{#2}}}
\newcommand{\sd}[2]{\frac{d^2{#1}}{d{#2}^2}}

 {\catcode`\|=\active \gdef|{\egroup\,\vrule\,\bgroup}}

\begin{document}
\title{Translation of an article by Paul Drude in 1904}
\author{Translation by 
A.~J.~Sederberg$^{1,}$\footnote{Present address: Division of Biological Sciences, University of Chicago, 5812 South Ellis Avenue, Chicago, IL 60637, USA.}~~(pp.~512--519, 560--561),\\ J.~Burkhart$^{2,}$\footnote{Present address: Department of Microsystems and Informatics, Hochschule Kaiserslautern, Amerikastrasse 1, D-66482 Zweibr\"{u}cken, Germany.}~~(pp.~519--527), and F.~Apfelbeck$^{3,}$\footnote{Present address: Faculty of Physics, Ludwig Maximilian University of Munich, Schellingstrasse 4, D-80799 Munich, Germany.}~~(pp.~528--560)\\ 
Discussion by B.~H.~McGuyer$^{1,}$\footnote{Present address: Department of Physics, Columbia University, 538 West 120th Street, New York, NY 10027-5255, USA.}\\
$^1$\normalsize{\textit{Department of Physics, Princeton University, Princeton, New Jersey 08544, USA}}\\
$^2$\normalsize{\textit{Department of Physics, Columbia University, 538 West 120th Street, New York, NY 10027-5255, USA}}}
\date{\today}
\maketitle			

\begin{abstract}
English translation of P.~Drude, Annalen der Physik 13, 512 (1904), an article by Paul Drude about Tesla transformers and wireless telegraphy.  
Includes a discussion of the derivation of an equivalent circuit and the prediction of nonreciprocal mutual inductance for Tesla transformers in the article, which is supplementary material for B.~McGuyer, PLoS ONE 9, e115397 (2014). 
\end{abstract}

\tableofcontents

\section{Introduction}
The German physicist Paul Drude (1863--1906)~\cite{hoffmann:2006} published a few articles~\cite{drude:1902, drude:1902oct, drude:1904, drude:1905} on the physics of Tesla transformers (or Tesla coils) in the beginning of the 20th century, during the era of wireless telegraphy (or early radio).  
These articles are of historical interest to the modeling of solenoids and Tesla transformers. 
In particular, Drude's 1904 article~\cite{drude:1904} is still cited as an important reference to justify the conventional equivalent circuit (or lumped-element model) for a Tesla transformer~\cite{skeldon:1997, denicolai:2002}.  
Unfortunately, no official English translation of this article exists.  
A partial translation covering pp.~512--514 \& 560--561 is available online \cite{antonioDrude}.  

This document presents an unofficial translation of Drude~\cite{drude:1904}. 
The equation numbering has been kept the same, and approximate page markings are included for reference.  
Drude~\cite{drude:1904} begins with a derivation of an equivalent circuit for a half-wave Tesla transformer (pp.~512--519), proceeds to solve the equations and analyze wireless telegraphy applications, and concludes with a summary of results (pp.~560--561).  
For reference, other discussions on modeling Tesla transformers are available in~\cite{skeldon:1997, denicolai:2002, denicolai:2001, boynton:1898, fleming:1919, terman:1943, smythe:1950}.  

During this translation, we were surprised to find Drude's prediction that \emph{the mutual inductance in the equivalent circuit for a Tesla transformer should be nonreciprocal} ($M_{12} \neq M_{21}$)!  
This mostly forgotten prediction is discussed in a multi-edition book on wireless telegraphy by J.~A.~Fleming~\cite{fleming:1919}, and is mentioned in the  books on inductance coils by E.~T.~Jones~\cite{jones:1921, jones:1932}.  
It is likely mentioned in other books from the wireless-telegraphy era, where Tesla transformers may be called ``oscillation transformers'' or ``Thomson coils'' (after Elihu Thomson).  
Relatedly, Hund~\cite{hund1st} treats mutual inductances in general as nonreciprocal, though does not mention Drude~\cite{drude:1904}. 
A modern treatment of Drude's nonreciprocity is available in McGuyer~\cite{mcguyer:2014}. 

This document ends with a discussion of the derivation of the equivalent circuit in Drude~\cite{drude:1904}.  
Some errors seem to have prevented the completion of this derivation, and lead to a different equivalent inductance for a resonant solenoid than that of Drude's 1902 article~\cite{drude:1902}.
We present a revised derivation which resolves this disagreement, and which results in nonreciprocities that agree with the treatment in McGuyer~\cite{mcguyer:2014}.

\section{Bibliographic information}
\begin{description} \setlength{\itemsep}{1pt} \setlength{\parskip}{0pt} \setlength{\parsep}{0pt}	
\item[Author:] Paul Karl Ludwig Drude
\item[Title:] ``\"{U}ber induktive Erregung zweier elektrischer Schwingungskreise mit Anwendung auf Periodenund D\"{a}mpfungsmessung, Teslatransformatoren und drahtlose Telegraphie''
\item[Translated Title:] ``Of inductive excitation of two electric resonant circuits with application to measurement of oscillation periods and damping, Tesla coils, and wireless telegraphy''
\item[Journal:] Annalen der Physik (abbreviated ``Ann.~Phys.'' or ``Ann.~d.~Phys.'')
\item[Volume:] 13
\item[Issue:] 3
\item[Pages:] 512--561
\item[Month:] February
\item[Year:] 1904
\item[Language:]  German
\item[IDS Number:] V24GL
\item[ISSN:] 0003-3804
\item[Online Volume:] 318
\item[Online ISSN:] 1521-3889
\item[DOI:] 10.1002/andp.18943180306
\item[URL:]  \url{http://dx.doi.org/10.1002/andp.18943180306}
\item[URL:]  \url{http://archive.org/stream/annalenderphysi103unkngoog#page/n599/mode/2up}
\item[Copyright:]   Wiley-VCH Verlag GmbH \& Co.~KGaA.  Reproduced with permission.
\end{description}

\newpage 

\section{Translation}
\begin{center}	
\rule[0.3em]{0.3 \columnwidth}{1.0pt} 
{\bf Start of article and translation}
\rule[0.3em]{0.3 \columnwidth}{1.0pt}\\
\rule[0.3em]{0.3 \columnwidth}{0.5pt} 
{\bf Page 512}
\rule[0.3em]{0.3 \columnwidth}{0.5pt}
\end{center}


\begin{center}
\textbf{5. Of inductive excitation of two electric resonant circuits with application to measurement of oscillation periods and damping, Tesla coils, and wireless telegraphy; \\ by P. Drude.}
\end{center}

Contents:  Introduction. I.~Definition and integration of the differential equations p.~513. II.~The magnetic coupling is very small p.~521. III.~Measuring the period and the damping p.~525. 1.~The maximum amplitude p.~528.  2.~The integral effect p.~530. IV.~The magnetic coupling is not very small p.~534. V.~The effectiveness of the Tesla transformer p.~540. VI.~Dependence of the Tesla effect on damping and coupling p.~544. VII.~Application to wireless telegraphy p.~550.  a) Simple or loosely-coupled receiver p.~551.  b) Tightly-coupled receiver p.~554. Main results p.~560.


J.~v.~Geitler\footnote{J.~v.~Geitler, Sitzungsber.~d.~k.~Akad.~d.~Wissensch.~zu Wien, Februar u.~Oktober 1895.}, B.~Galitzin\footnote{F\"{u}rst B.~Galitzin, Petersb.~Ber., Mai u.~Juni 1895.}, A.~Oberbeck\footnote{A.~Oberbeck, Wied.~Ann.~{\bf 55.}~p.~623.~1895.} and Domalip and Kol\'a\u{c}ek\footnote{R.~Domalip u.~F.~Kol\'a\u{c}ek, Wied.~Ann.~{\bf 57.}~p.~731.~1896.} have proven that, if two electric oscillating circuits interact strongly enough, each no longer has only one, but two, natural frequencies; this remains true even when the two systems are attuned with each other, that is to say, if they share a natural frequency and have weak or no interaction. This problem was later handled by M.~Wien\footnote{M.~Wien, Wied.~Ann.~{\bf 61.}~p.~151.~1897.} in a general and complete manner. Wien\footnote{M.~Wien, Ann.~d.~Phys.~{\bf 8.}~p.~686.~1902.} also applies his results to wireless telegraphy after the Braun System. V.~Bjerknes\footnote{V.~Bjerknes, Wied.~Ann.~{\bf 55.}~p.~120.~1895.} worked out in detail the case of very weak coupling, considering oscillation frequency and damping measurements through reference to the so-called resonance curve. 

\begin{center}	
\rule[0.3em]{0.3 \columnwidth}{0.5pt}
{\bf Page 513} 
\rule[0.3em]{0.3 \columnwidth}{0.5pt}
\end{center}

The treatment of this problem here differs from the aforementioned work in the following ways: 
\begin{enumerate} \setlength{\itemsep}{1pt} \setlength{\parskip}{0pt} \setlength{\parsep}{0pt}
\item The solution of the differential equations, especially the calculations of the amplitudes from the initial conditions, will be presented in a mathematically transparent way such that miscalculations are easily avoided, and which remains valid for more complicated applications, such as a strongly coupled transmitter and receiver in wireless telegraphy.\footnote{Addendum: I see that even Domalip and Kol\'a\u{c}ek have chosen a very similar treatment to this.}
\item The induced circuit will not, as in the cited work, be treated simply as a constant current-carrying wire, with a capacitor on the end, but rather in accordance with the actual conditions. 
\item Contrary to Wien's work, a difference appears in the results regarding the damping of both natural frequencies in a strongly coupled system.\footnote{It follows for the constants in the differential equations on p.~518 the conclusion $L_{12} < L_{21}$, whereas $L_{12} = L_{21}$ was otherwise assumed.}
\item Bjerknes only discusses the resonance curve of integral effects in detail. Here we shall also consult the resonance curve for maximum amplitude and thereby propose a simple experimental method for the determination of the individual attenuation of both oscillating circuits. 
\item The question of how best to construct a Tesla coil will be further addressed. Its resolution requires further experimental study still. 
\end{enumerate}

\subsection*{I.~Definition and Integration of the Differential Equations}\addcontentsline{toc}{subsubsection}{I.~Definition and Integration of the Differential Equations}
We initially assume that the secondary coil (e.g.,~a Tesla coil) lies centered and symmetric to the primary circuit. The general results also apply to any orientation of the secondary coil. The primary circuit includes the (electromagnetically measured) capacitance $C_1$. Let the potential difference between the plates of the capacitor $C_1$ at time $t$ be $V_1$, and let the current strength, that we can assume constant throughout the

\begin{center}	
\rule[0.3em]{0.3 \columnwidth}{0.5pt}
{\bf Page 514} 
\rule[0.3em]{0.3 \columnwidth}{0.5pt}
\end{center}

\noindent
entire primary circuit (since $C_1$ is chosen to be very large), be $i_1$. Let the number of magnetic field lines, which at any time loop around the primary circuit, be $N_1$. Then it follows:
\begin{align}
i_1 & = - C_1 \td{V_1}{t}, \label{eqn:1} \\
\sd{N_1}{t} + \; &w_1 \td{i_1}{t} + \frac{i_1}{C_1} = 0. \label{eqn:2}
\end{align}

$w_1$ is a coefficient on which the damping of the primary circuit depends (resistance from the wire and sparks, as well as the [minor] loss to radiation of the capacitor, eventually also the absorption of electrons in its dielectric). $w_1$ will also be assumed constant in time. Should this assumption not be met, then during the oscillation $w_1$ may be understood as the average. $N_1$ depends on $i_1$ and the current strength $i_2$ in the second coil, which is not constant along the coil. 

If we choose the axis of the coil (as if wrapped around a cylinder)  to be the $z$-direction, $z= 0$ lies in the middle of the coil, while both ends of the coil lie at $z = \pm h/2$ (so that $h$ is the height of the coil), then we can write for $i_2$ the first element of the Fourier series (fundamental mode):
\begin{align}
i_2 & = i^0_2 \cos \left( \frac{\pi z}{h} \right). \label{eqn:3}
\end{align}

$i^0_2$ is the current strength in the middle of the coil. In doing this so it is implied that the coil ends are free, without applied capacitance, so that it must be that $i_2 = 0$ for $z = \pm h/2$. 

The field lines $N_1$ now fall into two parts:
\begin{align} \label{eqn:4} 
N_1 & = N_{11} + N_{12}, 
\end{align}
of which the first part shall denote the field lines which only enclose the primary circuit, while $N_{12}$ shall count the field lines  which go around the primary circuit as well as the inductor coil.~(See Fig.~\ref{fig1}.)

For all the $N_{11}$ field lines the magnetomotive strength is the same, namely $4\pi i_1$, if circuit 1 can be considered linearly, as we wish to assume and as is adequately seen in practice.  Then $W_{11}$ is the magnetic 

\begin{center}	
\rule[0.3em]{0.3 \columnwidth}{0.5pt}
{\bf Page 515} 
\rule[0.3em]{0.3 \columnwidth}{0.5pt}
\end{center}

\noindent
resistance of the (connected in parallel) tube strength $N_{11}$, which is:\footnote{See for example P.~Drude, Physik d.~\"{A}thers p.~72.~Stuttgart 1894.}
\begin{align} \label{eqn:5}
N_{11} & = \frac{4 \pi i_1}{W_{11}}.
\end{align} 

For the field lines $N_{12}$, the magnetomotive strength is not uniform, since the coil cannot be treated as a

\begin{figure}[h]	
\begin{center}	
	\includegraphics[width=0.6\columnwidth]{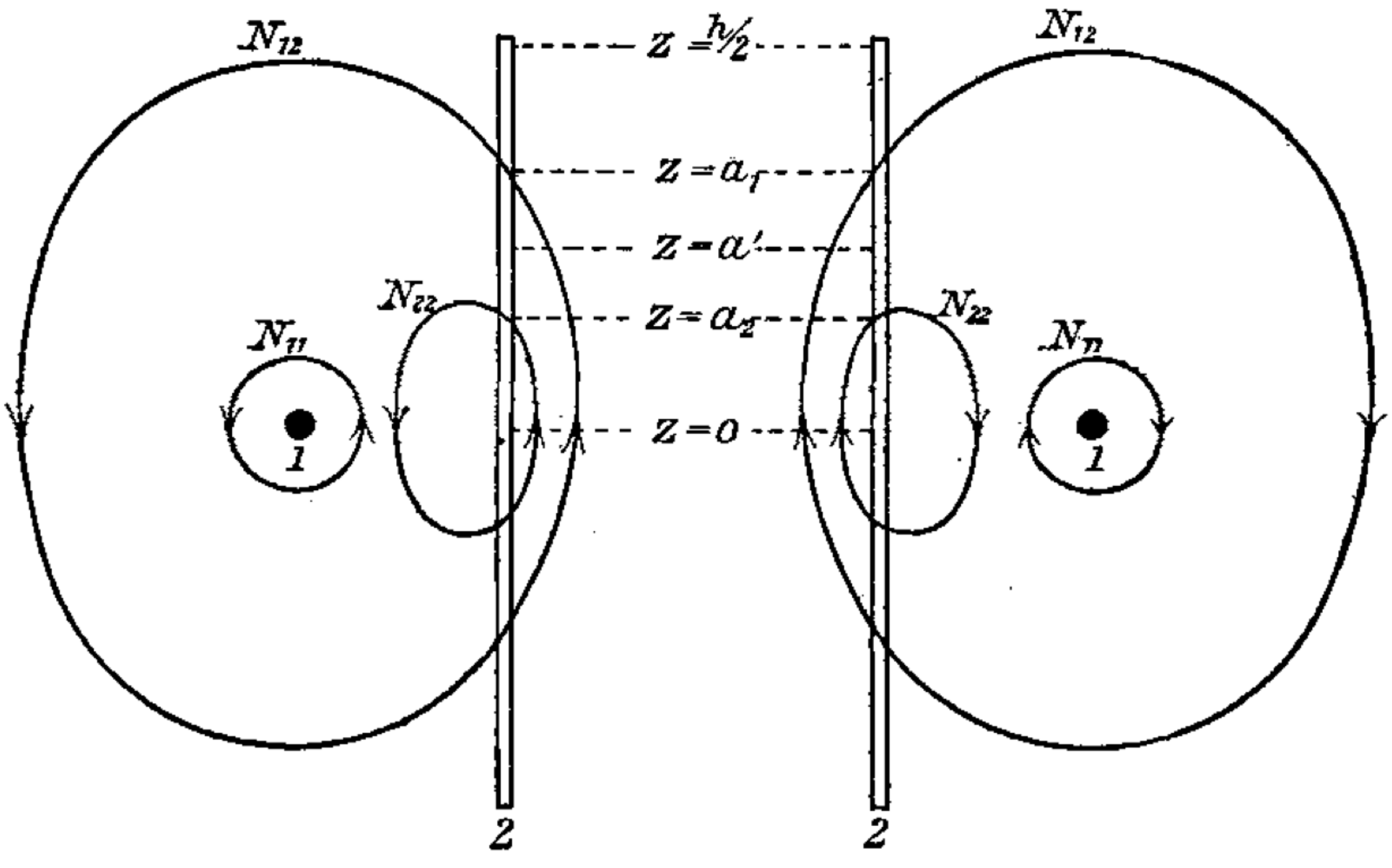}
	\caption{\label{fig1}Reproduction of Fig.~1 on p.~515 of Drude~\cite{drude:1904}. Copyright Wiley-VCH Verlag GmbH \& Co.~KGaA. Reproduced with permission.  } \addcontentsline{toc}{subsubsection}{Figure 1}
\end{center}
\end{figure}

\noindent
linear circuit. If one of these field lines cuts through the coil plane at position $z = \pm a$, then the magnetic force for this line is:\footnote{The $+$ sign before the second term on the right hand side applies if $i_1$ and $i_2$ are positive, which is expected in the same direction.}
\begin{align} \nonumber
4 \pi i_1 + 4 \pi \frac{n}{h} \int_{-a}^{+a} i_2 dz & = 4 \pi i_1 + 8 n i^0_2 \sin \left(\frac{\pi a}{h}\right), 
\end{align}
if $n$ denotes the total number of coil windings, so that along segment $dz$, there are $(n/h) dz$ windings. We 
can write an identical average electromotive force for the various field lines $N_{12}$:
\begin{align} \nonumber
4\pi i_1 + 8 n i_2^0 \sin \left(\pi \frac{a_1}{h} \right),
\end{align}
in which $a_1$ is smaller than $h/2$ and larger than $a'$, where the coil shell of the shortest field line that just barely loops around $i_1$ may be sliced. 

\begin{center}	
\rule[0.3em]{0.3 \columnwidth}{0.5pt}
{\bf Page 516} 
\rule[0.3em]{0.3 \columnwidth}{0.5pt}
\end{center}

It is therefore, according to the laws of magnetic circuits, written:
\begin{align}	\label{eqn:6}
N_{12} & = \frac{4 \pi i_1 + 8 n i^0_2 \sin (\pi a_1/h)}{W_{12}},
\end{align}
where $W_{12}$ denotes the magnetic resistance of all the tube strengths $N_{12}$ connected in parallel. 

From (\ref{eqn:2}) and (\ref{eqn:4}), it follows:
\begin{align}	\label{eqn:7}
4 \pi \left( \frac{1}{W_{11}} + \frac{1}{W_{12}} \right) \sd{i_1}{t} + \frac{8 n}{W_{12}} \sin \left( \frac{\pi a_1}{h} \right)
	\sd{i^0_2}{t} + w_1 \td{i_1}{t} + \frac{i_1}{C_1} & = 0.
\end{align}
	
The differential equation for the current strength $i_2$ in the coil includes $t$ and $z$ as independent variables. 

If $\mathfrak{e}_2$ denotes the electric charge (after electromagnetic measurements) along the horizontal length $dz=1$ of the coil, $\mathfrak{C}_2$ the capacitance of this length, and $V_2$ the potential of the coil at the position $z$, then it follows:
\begin{align}	\label{eqn:8}
\pd{i_2}{z} & = - \pd{\mathfrak{e}_2}{t}, \; \mathfrak{e}_2 = \mathfrak{C}_2 V_2.
\end{align}

Moreover, $N_2$ is the number of magnetic field lines that cross through the cross-section of the coil at height $z$, so it follows that between the positions $z$ and $z+dz$, between which $(n/h)dz$ windings lie, that  $i^2_2\mathfrak{w}_2 dz$ is the energy dissipation per time (from resistance and radiation):
\begin{align}	\label{eqn:9}
i_2 \mathfrak{w}_2 & = - \frac{n}{h} \pd{N_2}{t} - \pd{V_2}{z}.
\end{align}

From (\ref{eqn:8}) and (\ref{eqn:9}) it follows
\begin{align}	\label{eqn:10}
\frac{n}{h} \spd{N_2}{t} + \mathfrak{w}_2 \pd{i_2}{t} - \frac{1}{\mathfrak{C}_2} \spd{i_2}{z} & = 0.
\end{align}

We can rewrite this partial differential equation completely in terms of $i^0_2$, if we also write $N_2$ as a Fourier series, with the leading term remaining
\begin{align}	\label{eqn:11}
N_2 & = N^0_2 \cos \left( \frac{\pi z}{h} \right), 
\end{align}
and if we set $\mathfrak{w}_2$ and $\mathfrak{C}_2$, which depend simply on $z$,  to their average across the whole coil; that is, we treat them as constants. Then (\ref{eqn:3}), (\ref{eqn:10}), and (\ref{eqn:11}) yield
\begin{align}	\label{eqn:12}
\frac{n}{h} \spd{N^0_2}{t} + \mathfrak{w}_2 \td{i_2^0}{t} + \frac{\pi^2}{h^2 \mathfrak{C}_2} i^0_2 & = 0.
\end{align}

\begin{center}	
\rule[0.3em]{0.3 \columnwidth}{0.5pt}
{\bf Page 517} 
\rule[0.3em]{0.3 \columnwidth}{0.5pt}
\end{center}

From (\ref{eqn:11}), $N^0_2$ signifies the quantity of field lines that pass through the coil cross-section at the middle of the coil $z=0$. This quantity can also be separated into two parts
\begin{align}	\label{eqn:13}
N_2 & = N_{22} + N_{12},
\end{align}
where $N_{22}$ are the field lines that only pass through the coil, but do not go around circuit 1, while $N_{12}$ wind around the coil-windings as well as the primary circuit 1. For $N_{22}$ the magnetic force strength is no longer constant; we can however guess for it an average magnetic force strength
\begin{align}\nn
8 n i^0_2 \sin \left(\pi \frac{a_2}{h}\right)
\end{align}
where $a_2 < a'$, which is to say that $a_2 < a_1$ also. From this it follows that
\begin{align} \label{eqn:14}
N_{22} & = \frac{ 8 n i^0_2 \sin \left( \pi \frac{a_2}{h} \right)}{W_{22}} , 
\end{align}
if $W_{22}$ is the magnetic resistance of the $N_{22}$ tube strengths. In contrast, (\ref{eqn:12}) yields: 
\begin{align}	\label{eqn:15}
8n^2 \left( \frac{\sin \left(\pi\frac{a_1}{h}\right)}{W_{12}} + \frac{\sin \left(\pi \frac{a_2}{h}\right)}{W_{22}} \right) \sd{i^0_2}{t} + 
	\frac{4\pi n}{W_{12}} \sd{i_1}{t} + \mathfrak{w}_2 h \td{i^0_2}{t} + \frac{\pi^2}{\mathfrak{C}_2 h} i^0_2 & = 0.
\end{align}

This equation in combination with equation (\ref{eqn:7}) lays the framework for our problem. One can now write, for simplification, $i_2$ for $i_2^0$, that is to say \emph{from now on $i_2$ means the current strength in the middle winding of the Tesla coil}, so we have our framework in the well-known form:
\begin{align}	\label{eqn:16}
\left\{ \begin{aligned}
L_{11} \sd{i_1}{t} + L_{12} \sd{i_2}{t} + w_1 \td{i_1}{t} + \frac{i_1}{C_1} & = 0 , \\
L_{22} \sd{i_2}{t} + L_{21} \sd{i_1}{t} + w_2 \td{i_2}{t} + \frac{i_2}{C_2}  & = 0. 
\end{aligned} \right .
\end{align}

From there it follows:
\begin{align}	\label{eqn:17}
\left\{ \begin{aligned}
L_{11} & = 4 \pi \left(\frac{1}{W_{11}} + \frac{1}{W_{12}} \right), & L_{22} & = \frac{16 n^2}{\pi}
	 \left( \frac{\sin \left(\pi \frac{a_1}{h}\right)}{W_{11}} + \frac{\sin \left( \pi \frac{a_2}{h} \right)}{W_{22}} \right) , \\
L_{12} & = \frac{8 n}{W_{12}} \sin \left( \frac{\pi a_1}{h} \right) , & L_{21} & = \frac{8 n}{W_{12}} , \\ 
w_2 & = \frac{2}{\pi} \mathfrak{w}_2 h , & C_2 & = \mathfrak{C}_2 h:2\pi.
\end{aligned} \right .
\end{align}

\begin{center}	
\rule[0.3em]{0.3 \columnwidth}{0.5pt}
{\bf Page 518} 
\rule[0.3em]{0.3 \columnwidth}{0.5pt}
\end{center}

From (\ref{eqn:17}) it follows
\begin{align}	\label{eqn:18}
L_{12}:L_{21} = \sin \left( \frac{\pi a_1}{h} \right),
\end{align}
that is, \emph{we can no longer, as with two linear circuits, set $L_{12} = L_{21}$, but rather $L_{12}<L_{21}$}, and  even more so the smaller $a_1$ is; that is to say, the nearer to the primary circuit that the middle winding of the Tesla coil lies and the higher the Tesla coil is in relation to the primary circuit. \emph{At various positions of the Tesla coil relative to the primary circuit the relationship $L_{12}:L_{21}$ alternates, and it becomes even smaller, the stronger the mutual induction (magnetic coupling) between the Tesla coil and the primary circuit becomes.} This result, whose derivation came from the interpretation of the coefficients of the well-known equations (\ref{eqn:16}), \emph{also holds for any, including asymmetrical, positions of the Tesla coil relative to the primary circuit.} 

Of the coefficients $L$, $L_{11}$ and $L_{21}$ are relatively easy to calculate theoretically, while $L_{22}$ and $L_{12}$ may only be obtained after very tedious calculation. $L_{21}$ arises from the field lines, which current $i_i$ sends through the cross-sectional area $q$ of the Tesla coil. We call this count, if $i_1 = 1$, around any position $z$ of the coil $N_{21}$, so the first term $N^0_{21}$ of the Fourier series is 
\begin{align}\nn
N_{21} & = N^0_{21} \cos \left( \frac{\pi z}{h} \right)
\end{align}
given through
\begin{align} \label{eqn:19}
N_{21}^0 & = \frac{2}{h} \int_{-h/2}^{+h/2} N_{21} \cos \left( \frac{\pi z}{h} \right) dz.
\end{align}

Therefore, 
\begin{align}	\label{eqn:20}
L_{21} & = \frac{4n}{\pi h} \int_{-h/2}^{+h/2} N_{21} \cos \left( \frac{\pi z}{h} \right) dz.
\end{align}

If the primary windings are circuits, then the magnetic force originated through $i_1$ may be represented  by spherical harmonics at any point in space, and from there $N_{21}$, and thus also $L_{21}$, can be calculated. For the self-induction $L_{11}$, known formulas have already been worked out.\footnote{Stefan's formulas for the coils with oscillatory condenser discharges need a correction, see P.~Drude., Ann.~Phys.~{\bf 9.}~p.~604.~1902.}  
One must not use this formula for 

\begin{center}	
\rule[0.3em]{0.3 \columnwidth}{0.5pt}
{\bf Page 519} 
\rule[0.3em]{0.3 \columnwidth}{0.5pt}
\end{center}

\noindent
$L_{22}$, since $i_2$ is not constant along coil 2. 

From equation (\ref{eqn:16}), one can write equations for the potential difference $V_1$ across the capacitor $C_1$ in circuit 1, and for the potential $V_2$ (actually written as $V^h_2$) at a free end $z=h/2$ of the coil. With (\ref{eqn:1}) and (\ref{eqn:8}), it can be worked out from (\ref{eqn:16}) that:\footnote{These equations take the form of the equations by M. Wien (Ann.~d.~Phys.~{\bf 8.}~p.~694.~1902), if one reflects, that $2V_2$ is the potential difference between both coil ends. From there follows the valid equation 
\begin{align} \nonumber
i^0_2 = 2 C_2 \td{V^h_2}{t}.
\end{align}
The following developments are valid also for the case that the secondary circuit arises not from a coil, but from a linear circuit with a large added capacitor $C_2$. $2V_2$ is then the secondary potential difference.} 
\begin{align}	\label{eqn:21}
\left\{ \begin{aligned}
L_{11} C_1 \sd{V_1}{t} - 2 L_{12} C_2 \sd{V_2}{t} + w_1 C_1 \td{V_1}{t} + V_1 = 0, \\ 
L_{22} C_2 \sd{V_2}{t} - \frac{L_{21} C_1}{2} \sd{V_1}{t} + w_2 C_2 \td{V_2}{t} + V_2 = 0.
\end{aligned} \right.
\end{align}

We will rather use these equations in the form: 
\begin{align}	\tag{I}\label{eqn:I}
\left\{ \begin{aligned}
\frac{d^2 V_1}{dt^2} + 2 \delta_1 \frac{d V_1}{dt} + (\nu_1^2 + \delta_1^2) V_1 = p_{12} \frac{d^2 V_2}{dt^2}, \\
\frac{d^2 V_2}{dt^2} + 2 \delta_2 \frac{d V_1}{dt} + (\nu_2^2 + \delta_2^2) V_2 = p_{21} \frac{d^2 V_1}{dt^2}, 
\end{aligned} \right.
\end{align}
where it also follows: 
\begin{align}	\label{eqn:22}
\left\{ \begin{aligned}
\delta_1 &= \frac{w_1}{2 L_{11}}, & \nu_1^2 + \delta_1^2 &= \frac{1}{L_{11}C_1}, & p_{12} = 2 \frac{L_{12}C_2}{L_{11}C_1}, \\
\delta_2 &= \frac{w_2}{2 L_{22}}, & \nu_2^2 + \delta_2^2 &= \frac{1}{L_{22}C_2}, & p_{21} = \frac{1}{2} \frac{L_{21}C_1}{L_{22}C_2}, 
\end{aligned} \right.
\end{align}
or in the following form 
\begin{align}	\tag{II}\label{eqn:II}
\left\{ \begin{aligned}
(\tau_1^2 + \vartheta_1^2) \frac{d^2 V_1}{dt^2} + 2 \vartheta_1 \frac{d V_1}{dt} + V_1 = p_{12} (\tau_1^2 + \vartheta_1^2) \frac{d^2 V_2}{dt^2}, \\
(\tau_2^2 + \vartheta_2^2) \frac{d^2 V_2}{dt^2} + 2 \vartheta_2 \frac{d V_2}{dt} + V_2 = p_{21} (\tau_2^2 + \vartheta_2^2) \frac{d^2 V_1}{dt^2}, 
\end{aligned} \right.
\end{align}

\begin{center}	
\rule[0.3em]{0.3 \columnwidth}{0.5pt}
{\bf Page 520} 
\rule[0.3em]{0.3 \columnwidth}{0.5pt}
\end{center}

\noindent
where it follows 
\begin{align}	\label{eqn:23}
\left\{ \begin{aligned}
\vartheta_1 = \frac{1}{2} w_1 C_1, & & \tau_1^2+\vartheta_1^2 &= L_{11}C_1, & & p_{12} = 2 \frac{L_{12}C_2}{L_{11}C_1}, \\
\vartheta_2 = \frac{1}{2} w_2 C_2, & &\tau_2^2+\vartheta_2^2 &= L_{22}C_2, & & p_{21} = \frac{1}{2} \frac{L_{21}C_1}{L_{22}C_2}. 
\end{aligned} \right.
\end{align}

As an integral of the form (\ref{eqn:I}) we choose 
\begin{align}	\label{eqn:24}
\begin{aligned}
V_1 &= A e^{xt}, &V_2 &= B e^{xt}, & x &= -\delta + i \nu, 
\end{aligned}
\end{align}
as the integral of (\ref{eqn:II}) we choose: 
\begin{align}	\label{eqn:25}
\begin{aligned}
V_1 &= A e^{t/y}, &V_2 &= B e^{t/y}, & y &= -\vartheta + i \tau. 
\end{aligned}
\end{align}
Here $i = \sqrt{-1}$. 

It will be proven, that in one case form (\ref{eqn:I}) is practical and in the other case form (\ref{eqn:II}). 

From (\ref{eqn:I}) and (\ref{eqn:24}) it follows: 
\begin{align}	\label{eqn:26}
\left\{ \begin{aligned}
A(x^2 + 2 \delta_1 x + \nu_1^2 + \delta_1^2) &= p_{12} x^2 B, \\
B(x^2 + 2 \delta_2 x + \nu_2^2 + \delta_2^2) &= p_{21} x^2 A. 
\end{aligned} \right.
\end{align}

In contrast it follows from (\ref{eqn:II}) and (\ref{eqn:25}): 
\begin{align}	\label{eqn:27}
\left\{ \begin{aligned}
A(\tau_1^2 + \vartheta_1^2 + 2 \vartheta_1 y + y^2 ) &= p_{12} (\tau_1^2 + \vartheta_1^2) B, \\
B(\tau_2^2 + \vartheta_2^2 + 2 \vartheta_2 y + y^2 ) &= p_{21} (\tau_2^2 + \vartheta_2^2) A. 
\end{aligned} \right.
\end{align}

From elimination of $A$:$B$ it follows for $x$ and $y$ the biquadratic equations: 
\begin{align} 	\label{eqn:28}
(x^2 + 2 \delta_1 x + \nu_1^2 + \delta_1^2)(x^2 + 2 \delta_2 x + \nu_2^2 + \delta_2^2) = p_{12} p_{21} x^4, \\
\left\{ \begin{aligned}
(y^2 + 2 \vartheta_1 y + \tau_1^2 + \vartheta_1^2)(y^2 &+ 2 \vartheta_2 y + \tau_2^2 + \vartheta_2^2) \\
	&= p_{12} p_{21} (\tau_1^2 + \vartheta_1^2)(\tau_2^2 + \vartheta_2^2).  \label{eqn:29}
\end{aligned} \right.
\end{align}

If $p_{12}=0$, which means the primary circuit is present all alone, then (\ref{eqn:26}) results in: 
\begin{align}	\label{eqn:30}
x = - \delta_1 \pm i \nu_1. 
\end{align}
$\nu_1$ thus indicates the frequency, $\delta_1$ the damping of the natural oscillations of the primary circuit. If one names $T_1$ as the period of oscillation, $\gamma_1$ as the logarithmic decrement, it follows 
\begin{align}	\tag{30'}\label{eqn:30p}
\begin{aligned}
\nu_1 &= \frac{2 \pi}{T_1}, & \delta_1 &= \frac{\gamma_1}{T_1}. 
\end{aligned}
\end{align}
Simultaneously, $\nu_2$ and $\delta_2$ indicate the frequency respectively the damping of the natural oscillation of the Tesla coil on its own. $\vartheta_1, \tau_1, \vartheta_2, \tau_2$ have

\begin{center}	
\rule[0.3em]{0.3 \columnwidth}{0.5pt}
{\bf Page 521} 
\rule[0.3em]{0.3 \columnwidth}{0.5pt}
\end{center}

\noindent
no directly descriptive meanings, nevertheless the natural oscillations are the following: 
\begin{align}	\label{eqn:31}
\begin{aligned}
y &= -\vartheta_1 \pm i \tau_1 & &\text{respectively} & y &= - \vartheta_2 \pm i \tau_2.
\end{aligned}
\end{align}

If $\vartheta_1^2$ is insignificant compared to $\tau_1^2$, and $\vartheta_2^2$ compared to $\tau_2^2$, it follows that: 
\begin{align}	\nonumber
e^{\frac{t}{-\vartheta_1 + i \tau_1}} = e^{-\frac{\vartheta_1}{\tau_1^2}t} \cdot e^{-i \frac{t}{\tau_1}}. 
\end{align}
If one thus sets
\begin{align}	\nonumber
e^{\frac{t}{-\vartheta_1 + i \tau_1}} = e^{-\gamma \cdot \frac{t}{T_1}} \cdot e^{i \frac{2 \pi}{T_1} t}, 
\end{align}
if follows that 
\begin{align}	\label{eqn:32}
\begin{aligned}
\tau_1 &= \frac{T_1}{2\pi}, & \vartheta_1 &= \frac{\gamma_1 T_1}{4 \pi^2}. 
\end{aligned}
\end{align}
Here $T_1$ is the period, $\gamma_1$ the logarithmic decrement of the circuit 1. 

The equations (\ref{eqn:I}), (\ref{eqn:24}), and (\ref{eqn:28}) are now useful, if {\it the magnetic coupling} $k^2$ 
\begin{align}	\label{eqn:33}
k^2 = p_{12}p_{21} = \frac{L_{12}L_{21}}{L_{11}L_{22}} 
\end{align}
is so small, that it is insignificant compared to 1. 

Let us first consider this case. 

\begin{center}
\textbf{II. The Magnetic Coupling is Very Small.}\addcontentsline{toc}{subsubsection}{II.~The Magnetic Coupling is Very Small}
\end{center}

If $p_{12}p_{21}$ is insignificant compared to 1, (\ref{eqn:28}) results in the four roots of $x$: 
\begin{align}	\label{eqn:34}
\begin{aligned}
x_1 &= - \delta_1 + i \nu_1, & x_2 &= - \delta_1 - i \nu_1, & x_3 &= - \delta_2 + i \nu_2, & x_4 &= - \delta_2 - i \nu_2.  
\end{aligned}
\end{align}
Therefore the general integral of (\ref{eqn:I}) is: 
\begin{align} 	\label{eqn:35}
\left\{ \begin{aligned}
V_1 &= A_1 e^{x_1 t} + A_2 e^{x_2 t} + A_3 e^{x_3 t} + A_4 e^{x_4 t}, \\
V_2 &= B_1 e^{x_1 t} + B_2 e^{x_2 t} + B_3 e^{x_3 t} + B_4 e^{x_4 t}.  
\end{aligned} \right.
\end{align}

From (\ref{eqn:1}) follows: 
\begin{align}	\label{eqn:36}
-\frac{i_1}{C_1} = x_1 A_1 e^{x_1 t} + x_2 A_2 e^{x_2 t} + x_3 A_3 e^{x_3 t} + x_4 A_4 e^{x_4 t}, 
\end{align}
from (\ref{eqn:3}) and (\ref{eqn:8}) follows: 
\begin{align}	\label{eqn:37}
\pi \frac{i_2}{\mathfrak{C}_2 h} = x_1 B_1 e^{x_1 t} + x_2 B_2 e^{x_2 t} + x_3 B_3 e^{x_3 t} + x_4 B_4 e^{x_4 t}. 
\end{align}

The starting condition for $t=0$ is: 
\begin{align}	\label{eqn:38}
\begin{aligned}
V_1 &= F, & V_2 &= 0, & i_1 &= 0, & i_2 = 0. 
\end{aligned}
\end{align}


\begin{center}	
\rule[0.3em]{0.3 \columnwidth}{0.5pt}
{\bf Page 522} 
\rule[0.3em]{0.3 \columnwidth}{0.5pt}
\end{center}

Therefore (\ref{eqn:35}), (\ref{eqn:36}), (\ref{eqn:37}), and (\ref{eqn:38}) are obtaining: 
\begin{gather*}	\label{eqn:39}
\begin{align}	
\begin{aligned} 
\sum A_n &= F, & \sum A_n x_n &= 0, & \sum B_n &= 0, & \sum B_n x_n = 0,  
\end{aligned}
\end{align}\\
n = 1, 2, 3, 4. 
\end{gather*} 

Now the second of the two equations (\ref{eqn:26}) is: 
\begin{align}	\label{eqn:40} 
B_n \left(x_n + 2 \delta_2 + \frac{\nu_2^2 + \delta_2^2}{x_n} \right) = p_{21} x_n A_n. 
\end{align}

If we sum up these four established equations for $n$ = 1, 2, 3, 4  we get because of equation (\ref{eqn:39}): 
\begin{align}	\label{eqn:41} 
\frac{B_1}{x_1} + \frac{B_2}{x_2} + \frac{B_3}{x_3} + \frac{B_4}{x_4} = 0.
\end{align}
Therefore it follows form the second equation (\ref{eqn:26}): 
\begin{align}	\label{eqn:42} 
B_n \left(1 + \frac{2 \delta_2}{x_n} + \frac{\nu_2^2 + \delta_2^2}{x_n^2} \right) = p_{21} A_n. 
\end{align}

If we sum up these four established equations for $n$ = 1, 2, 3, 4 we get because of equations (\ref{eqn:39}) and (\ref{eqn:41}): 
\begin{align}	\label{eqn:43} 
\frac{B_1}{x_1^2} + \frac{B_2}{x_2^2} + \frac{B_3}{x_3^2} + \frac{B_4}{x_4^2} = p_{21} \frac{F}{\nu_2^2 + \delta_2^2}. 
\end{align}
Therefore (\ref{eqn:43}), (\ref{eqn:41}), and the last two equations of (\ref{eqn:39}) result in: 
\begin{align}	\label{eqn:44} 
\frac{B_1}{x_1^2} \cdot 
\begin{vmatrix}
1, & 1, & 1, & 1 \\
x_1, & x_2, & x_3, & x_4 \\
x_1^2, & x_2^2, & x_3^2, & x_4^2 \\
x_1^3, & x_2^3, & x_3^3, & x_4^3 
\end{vmatrix} 
= p_{21} \frac{F}{\nu_2^2 +\delta_2^2} x_2 x_3 x_4 \cdot 
\begin{vmatrix}
1, & 1, & 1 \\
x_2, & x_3, & x_4 \\
x_2^2, & x_3^2, & x_4^2
\end{vmatrix} 
\end{align}

The occurring determinants can be written down directly.  For it is 
\begin{align}	\label{eqn:45}  
\begin{vmatrix}
1, & 1, & 1, & 1 \\
x_1, & x_2, & x_3, & x_4 \\
x_1^2, & x_2^2, & x_3^2, & x_4^2 \\
x_1^3, & x_2^3, & x_3^3, & x_4^3 
\end{vmatrix} 
= + (x_1-x_2)(x_1-x_3)(x_1-x_4)(x_2-x_3)(x_2-x_4)(x_3-x_4). 
\end{align}
Therefore (\ref{eqn:44}) results in: 
\begin{align}	\label{eqn:46} 
B_1(x_1-x_2)(x_1-x_3)(x_1-x_4) = - p_{21} \frac{F}{\nu_2^2+\delta_2^2} x_1^2 x_2 x_3 x_4, 
\end{align}
and likewise it follows (cyclically swapping the letters): 
\begin{align}	\label{eqn:47} 
B_2(x_2-x_3)(x_2-x_4)(x_2-x_1) = - p_{21} \frac{F}{\nu_2^2+\delta_2^2} x_1 x_2^2 x_3 x_4, \\
B_3(x_3-x_4)(x_3-x_1)(x_3-x_2) = - p_{21} \frac{F}{\nu_2^2+\delta_2^2} x_1 x_2 x_3^2 x_4, \label{eqn:48} \\ 
B_4(x_4-x_1)(x_4-x_2)(x_4-x_3) = - p_{21} \frac{F}{\nu_2^2+\delta_2^2} x_1 x_2 x_3 x_4^2. \label{eqn:49} 
\end{align}

\begin{center}	
\rule[0.3em]{0.3 \columnwidth}{0.5pt}
{\bf Page 523} 
\rule[0.3em]{0.3 \columnwidth}{0.5pt}
\end{center}

The potential $V_2$ at the end of the coil is of special interest.  By (\ref{eqn:35}) and (\ref{eqn:46}) to (\ref{eqn:49}) the same meaning of the potential difference inside the primary circuit in the beginning is demonstrated. 

$B_2$ is the complex conjugate to $B_1$, as well as $B_4$ to $B_3$ and $x_2$ to $x_1$ and $x_4$ to $x_3$. 

Therefore $V_2$ is after (\ref{eqn:35}) equal to a real term. 

To keep the overview over the result, one can write equation (\ref{eqn:36}) as the following, if for the $x$ values (\ref{eqn:34}) is inserted: 
\begin{align} 	\label{eqn:50}
\left\{ \begin{aligned}
V_2 &= e^{i \frac{\nu_1+\nu_2}{2} t} \left[ B_1 e^{-\delta_1 t} e^{+i \frac{\nu_1-\nu_2}{2} t} + B_3 e^{-\delta_2 t} e^{-i \frac{\nu_1-\nu_2}{2} t} \right] \\ 
	&+ e^{-i \frac{\nu_1+\nu_2}{2} t} \left[ B_2 e^{-\delta_1 t} e^{-i \frac{\nu_1-\nu_2}{2} t} + B_4 e^{-\delta_2 t} e^{+i \frac{\nu_1-\nu_2}{2} t} \right]. 
\end{aligned} \right.
\end{align}
If the following is valid 
\begin{align}	\tag{50'}\label{eqn:50p}
B_1 e^{-\delta_1 t} e^{+i \frac{\nu_1-\nu_2}{2} t} + B_3 e^{-\delta_2 t} e^{-i \frac{\nu_1-\nu_2}{2} t} = - i B e^{i \chi}, 
\end{align}
where $B$ and $\chi$ are real terms, it follows as 
\begin{align}	\label{eqn:51} 
V_2 = 2 B \sin \left( \frac{\nu_1 + \nu_2}{2} t + \chi \right). 
\end{align}
$B$ {\it thus names the amplitude of the potential at the end of the Tesla coil.} There are emerging beats for $V_2$, as $B$ depends on time. 

From (\ref{eqn:50}) it follows, as $B_2$ is the complex conjugate of $B_1$, and $B_4$ to $B_3$: 
\begin{align} 	\label{eqn:52}
\left\{ \begin{aligned}
B^2 &= B_1 B_2 e^{-2 \delta_1 t} + B_3 B_4 e^{-2 \delta_2 t} \\ 
	&+ e^{-(\delta_1 + \delta_2) t} \left[ B_1 B_4 e^{i(\nu_1 - \nu_2) t} + B_3 B_2 e^{-i(\nu_1-\nu_2)t} \right]. 
\end{aligned} \right.
\end{align}
Now it follows from (\ref{eqn:46}) 
\begin{align}	\label{eqn:53} 
B_1 = - p_{21} F (\delta_1^2 + \nu_1^2) \frac{-\delta_1 + i \nu_1}{2 i \nu_1 \substack{[-(\delta_1-\delta_2) + i (\nu_1-\nu_2)] \\ [-(\delta_1-\delta_2) + i (\nu_1+\nu_2)]}}
\end{align}
As $\delta_1-\delta_2$ is always small compared to $\nu_1 + \nu_2$, and as well $\delta_1$ compared to $\nu_1$, it is possible to set until the first order in $\delta_1$ respectively $\delta_2$: 
\begin{align}	\label{eqn:54} 
B_1 = p_{21} F \frac{\nu_1^2}{2(\nu_1+\nu_2)}\left(1 + i \frac{\delta_1\nu_2 + \delta_2\nu_1}{\nu_1(\nu_1+\nu_2)}\right) \cdot \frac{1}{\nu_1-\nu_2+i(\delta_1-\delta_2)}. 
\end{align}

\begin{center}	
\rule[0.3em]{0.3 \columnwidth}{0.5pt}
{\bf Page 524} 
\rule[0.3em]{0.3 \columnwidth}{0.5pt}
\end{center}

Simultaneously it follows 
\begin{align} 	\tag{54}
\left\{ \begin{aligned}
B_2 &= \hphantom{-}p_{21} F\cdot \frac{\nu_1^2}{2(\nu_1+\nu_2)} \left( 1 - i \frac{\delta_1 \nu_2 + \delta_2 \nu_1}{\nu_1(\nu_1+\nu_2)} \right) \cdot \frac{1}{\nu_1-\nu_2-i(\delta_1-\delta_2)}, \\
B_3 &= -p_{21} F\cdot \frac{\nu_1^2}{2(\nu_1+\nu_2)} \left( 1 + i \frac{\delta_1 \nu_2 + \delta_2 \nu_1}{\nu_1(\nu_1+\nu_2)} \right) \cdot \frac{1}{\nu_1-\nu_2-i(\delta_1-\delta_2)}, \\
B_4 &= -p_{21} F\cdot \frac{\nu_1^2}{2(\nu_1+\nu_2)} \left( 1 - i \frac{\delta_1 \nu_2 + \delta_2 \nu_1}{\nu_1(\nu_1+\nu_2)} \right) \cdot \frac{1}{\nu_1-\nu_2-i(\delta_1-\delta_2)}. 
\end{aligned} \right.
\end{align}
Therefore (\ref{eqn:52}) results in: 
\begin{align}	\label{eqn:55} 
B = p_{21} F \frac{\nu_1^2}{2(\nu_1+\nu_2)} \sqrt{\frac{e^{-2\delta_1 t} + e^{-2 \delta_2 t} - 2 e^{-(\delta_1+\delta_2)t} \cos \varphi}{(\nu_1-\nu_2)^2+(\delta_1+\delta_2)^2}}, 
\end{align}
with 
\begin{align}	\label{eqn:56} 
\varphi = (\nu_1-\nu_2)t + \frac{\delta_1\nu_2+\delta_2\nu_1}{\nu_1+\nu_2}\left(\frac{1}{\nu_1}-\frac{1}{\nu_2}\right). 
\end{align}
It is also possible to write $B$ in the following form: 
\begin{align}	\label{eqn:57} 
B = p_{21} F \frac{\nu_1^2}{2(\nu_1+\nu_2)} e^{- \frac{\delta_1+\delta_2}{2} t} \sqrt{\frac{\left( e^{+\frac{\delta_1-\delta_2}{2} t} - e^{-\frac{\delta_1-\delta_2}{2} t} \right)^2 + 4 \sin^2 \frac{1}{2} \varphi}{(\nu_1-\nu_2)^2+(\delta_1+\delta_2)^2}}. 
\end{align}

For the case of resonance, $\nu_1=\nu_2$ results in 
\begin{align}	\label{eqn:58} 
B = p_{21} F \frac{\nu_1}{4} \frac{e^{-\delta_1 t} - e^{-\delta_2 t}}{\delta_1-\delta_2}. 
\end{align}
$B$ raises to a maximum value, if 
\begin{align}	\label{eqn:59} 
\begin{aligned}
e^{(\delta_1-\delta_2)t} &= \delta_1:\delta_2, & &{\rm i.e.~for} & t &= \frac{{\rm log}\delta_1/\delta_2}{\delta_1-\delta_2}. 
\end{aligned}
\end{align}
This maximum value (for the case of resonance) is according to (\ref{eqn:58}) and (\ref{eqn:59}): 
\begin{align} 	\label{eqn:60}
\left\{ \begin{aligned}
B_\text{Max.} &= - p_{21} F \frac{\nu_1}{4 \delta_2} \cdot \left( \frac{\delta_1}{\delta_2} \right)^{\frac{\delta_1}{\delta_2-\delta_1}} \\
	&= - p_{21} F \frac{\nu_1}{4} \frac{\delta_1^{\delta_1/\delta_2-\delta_1}}{\delta_2^{\delta_2/\delta_2-\delta_1}}. 
\end{aligned} \right.
\end{align}

This coincides with the formula from Bjerknes\footnote{V.~Bjerknes, Wied.~Ann.~{\bf 55.}~p.~134, Formel (9a). 1895.}

For $\nu_1 = \nu_2$, $\delta_1=\delta_2$ (\ref{eqn:58}) results in the indefinite form 0/0. However, if one develops in (\ref{eqn:57}) after the powers of $\nu_1-\nu_2$ and $\delta_1-\delta_2$, there arises, if one replaces $\varphi$ in (\ref{eqn:56}) with $\varphi = (\nu_1-\nu_2)t$, a certain form, as $(\nu_1-\nu_2)^2+(\delta_1 - \delta_2)^2$ is occurring in the numerator and denominator. Therefore it follows: 
\begin{align}	\label{eqn:61} 
B = p_{21} F \frac{\nu_1}{4} t e^{-\delta_1 t} , 
\end{align}

\begin{center}	
\rule[0.3em]{0.3 \columnwidth}{0.5pt}
{\bf Page 525} 
\rule[0.3em]{0.3 \columnwidth}{0.5pt}
\end{center}

\noindent
which is also derived from Bjerknes.\footnote{l.~c., Formel (8a). From this follows: 
\begin{align}\nonumber
B_\text{Max} = p_{21} F \frac{\nu_1}{4 e \delta_1}. 
\end{align}}  
Whether it is allowed for $\varphi$, to conduct the border transition $\nu_1=\nu_2$ earlier in its second element, as in its first element $(\nu_1-\nu_2)t$, should be investigated in the following.\footnote{The case $\nu_1=\nu_2$, $\delta_1=\delta_2$ demands technically speaking a special treatment beginning from (\ref{eqn:28}) already, as it says then: $x^2+2 \delta_1 x + \nu_1^2 + \delta_1^2 = \pm x^2 k$, and $k$ cannot be disregarded against 1 anymore. For this reason, it is impossible, that the roots of $x$ get exactly equal pairwise, what seems to arise from (\ref{eqn:34}) for $\delta_1=\delta_2$, $\nu_1=\nu_2$.} 

\begin{center}
\textbf{III. Measurement of the Period and the Damping.}\addcontentsline{toc}{subsubsection}{III.~Measurement of the Period and the Damping}
\end{center}

We assume the secondary circuit is unchangeable. Of practical interest is both the question, whether the effect in the secondary circuit forms a maximum under the continuous change of the frequency $\nu_1$ of the primary circuit when exactly $\nu_1 = \nu_2$, as well as the question, whether one can determine the dampings $\delta_1$ and $\delta_2$ of the two oscillating circuits, only from quantitative measurement of the resonance curve, which means from the quantitative measurement of the effect in the secondary circuit with a different frequency $\nu_1$ of the primary circuit.
Bjerknes discussed both questions, but only sufficing for the integral effect, which is authoritative for electrometric, bolometric, thermoelectric (Joule heat) measurement methods. Insofar as the method needs to be generalized, it makes no difference, whether one changes the capacity $C_1$ to change $\nu_1$, while $L_{11}$ stays at the same value\footnote{This case is based upon my observations (Ann.~d.~Phys.~{\bf 9.}~p.~293. 1902).} (case a), or one contrary keeps $C_1$ at the same value, but varies $L_{11}$ (case b). 
In case a) the effect in the secondary circuits is that the decisive coefficient $p_{21}$ is not constant in (\ref{eqn:22}); 
it is, if $\delta^2$ is neglected against $\nu^2$, that is, $\gamma^2$ is neglected against $4 \pi^2$, to be converted in the following form:
\begin{align}	\label{eqn:62} 
p_{21} = \frac{1}{2} \frac{L_{21}}{L_{11}} \frac{\nu_2^2}{\nu_1^2} ;
\end{align}

\begin{center}	
\rule[0.3em]{0.3 \columnwidth}{0.5pt}
{\bf Page 526} 
\rule[0.3em]{0.3 \columnwidth}{0.5pt}
\end{center}

\noindent
On the contrary, in case b) $p_{21}$ is constant, assuming that the coefficient of the mutual induction $L_{21}$ stays constant upon the change of $L_{11}$, which can be easily realized experimentally. The observations of Bjerknes are based only on the case b). 
But also in other respects an extension of this fundamental work is desirable, namely to treat the maximum amplitude also exhaustively, 
because it turns out that by combining the observations of the resonance curve of the integral effect and the maximum amplitude, the dampings $\delta_1$ and $\delta_2$ of both circuits may be calculated separately. This is in my option an easier observation method than the one proposed by Bjerknes. The maximum amplitude (maximum value of the potential $V_2$) may be obtained quantitatively by experiment, through sparking distance, or better by the electrical distraction of the cathode beams in a Braun tube.

The observations both for the integral and for the maximum effect tie in with formula (\ref{eqn:55}). If one sets 
\begin{align}	\label{eqn:63} 
\nu_1 = \nu_2 (1 + \zeta), 
\end{align}
then in the secondary circuit the effect $E$ is more generally 
\begin{align}	\label{eqn:64} 
E = P(1 + a \zeta - b \zeta^2), 
\end{align}
where $a$ may be positive or negative, but $b$ always stays positive. $b$ is very large compared to 1. The maximum $E_m$ of $E$ occurs by (\ref{eqn:64}) as 
\begin{align}	\label{eqn:65} 
\zeta_m = \frac{a}{2b} 
\end{align}
and has the value of 
\begin{align}	\label{eqn:66} 
E_m = P\left(1 + \frac{a^2}{4b} \right). 
\end{align}
If one sets 
\begin{align}	\label{eqn:67} 
\begin{aligned} 
\zeta &= \zeta_m + \eta, & \nu_1 &= \nu_2(1 + \zeta_m + \eta), 
\end{aligned} 
\end{align}
where $\eta$ describes the percent difference of the frequency $\nu_1$ from the {\it resonant frequency} $\nu_2(1+\zeta_m)$, which 

\begin{center}	
\rule[0.3em]{0.3 \columnwidth}{0.5pt}
{\bf Page 527} 
\rule[0.3em]{0.3 \columnwidth}{0.5pt}
\end{center}

\noindent
describes the frequency that has the major effect in the secondary circuit, it follows from (\ref{eqn:64}): 
\begin{align}	\label{eqn:68} 
E = P\left( 1 + \frac{a^2}{4b} - b \eta^2 \right), 
\end{align}
that means according to (\ref{eqn:66}): 
\begin{align}	\label{eqn:69} 
\frac{E_m-E}{E_m} = b \eta^2. 
\end{align}
$\eta$ can be determined experimentally, if the primary circuit is calibrated onto the oscillation period or the wavelength. Hence, (\ref{eqn:69}) states, in which way it is possible to calculate the coefficient $b$ quantitative from the resonance curve. The resonance curve is even steeper, the bigger $b$ is. If $b$ is found, it is needed to calculate the oscillation period $\nu_2$ from (\ref{eqn:65}) and (\ref{eqn:67}), if one knows $a$.

We will now get to know the more particular examination, that it is, because of the dependence\footnote{The dependence may persist both because of a changed frequency, as well as a changed current $i_1$ (with varied $C_1$). Bjerknes also takes this into consideration.} of the damping $\delta_1$ from the frequency $\nu_1$, not possible to state the coefficient $a$ arithmetically, but definitely to derive $b$ from (\ref{eqn:55}) theoretically. $b$ does not only depend on the logarithmical decrements $\gamma_1$ and $\gamma_2$ (respectably the dampings $\delta_1$ and $\delta_2$) of the two oscillation circuits. {\it One therefore gets information from the resonance curves about both dampings depending on the dimension $b$, while not\footnote{
The seemingly differing result of Bjerknes is only caused since he expands the resonance curve to higher values $\gamma$ than we want to do it here. In fact, Bjerkenes' determination of the difference of the isochronity point and the resonance point (the parameter Bjerknes calls $S$) is also only an estimate of the order of magnitude. Compare: V.~Bjerknes, l.~c.~p.~150.} about the accurate value of the frequency $\nu_2$}. However, if one combines the resonance observations of the two different effects (maximum and integral value), one can determine $\nu_2$ accurately too, as well as the dampings of both oscillation circuits individually. This shall be shown now.

We discuss the individual cases now separately. 


\begin{center}	
\rule[0.3em]{0.3 \columnwidth}{0.5pt}
{\bf Page 528} 
\rule[0.3em]{0.3 \columnwidth}{0.5pt}
\end{center}

1.~{\it The maximum amplitude.} 
\textit{a) $C_1$ is changed.} Formulas (\ref{eqn:57}) and (\ref{eqn:62}) yield for the highest amplitude achieved over time of the potential $V_2$, with the value\footnote{Technically speaking, this value of $t$ yields the biggest amplitude of $B_\text{Max}$ only for $\nu_1=\nu_2$. Since $\nu_1$ is assumed to be close to $\nu_2$ we can use this value of $t$ though. The consideration of this change of $t$ would have only influence on the coefficient $a$ in formula (\ref{eqn:64}), which is not crucial, following the considerations in the text above.}) of $t$ following from (\ref{eqn:59}):
\begin{align}	\label{eqn:70}
B_{\rm Max.} = F \frac{L_{21}}{L_{11}} \cdot \frac{\nu_2^2}{4(\nu_1+\nu_2)}\cdot q \sqrt{\frac{1 + \delta_1 \delta_2 \left( \frac{{\rm log}\delta_1/\delta_2}{\delta_1-\delta_2}\right)^2 \left( \frac{\nu_1-\nu_2}{\delta_1-\delta_2}\right)^2}{1 + \left( \frac{\nu_1-\nu_2}{\delta_1-\delta_2}\right)^2}}, 
\end{align}
while there is
\begin{align}	\label{eqn:71}
q = \frac{\delta_2^{\delta_2/\delta_1-\delta_2}}{\delta_1^{\delta_1/\delta_1-\delta_2}}. 
\end{align}

Introducing (\ref{eqn:63}), it is:
\begin{align}	\label{eqn:72}
B_{\rm Max} = F \frac{L_{21}}{L_{11}} \cdot \frac{\nu_2}{8(1 + \frac{1}{2} \zeta)} \left( q + \frac{\partial q}{\partial \delta_1} \cdot \frac{d \delta_1}{d \zeta}\right) \sqrt{1-b\zeta^2}, 
\end{align}
where it is 
\begin{align}	\label{eqn:73}
b = \left(\frac{\nu_2}{\delta_1-\delta_2}\right)^2\left( 1 - \delta_1 \delta_2 \left( \frac{{\rm log}\delta_1/\delta_2}{\delta_1-\delta_2}\right)^2 \right). 
\end{align}

This quantity $b$ is continuously positive and very large against 1. 
Indeed, a series expansion yields to:
\begin{align}	\label{eqn:74}
b &= \frac{\nu_2^2}{12 \delta_1^2} \left( 1 + x + \frac{13}{15} x^2 + \frac{11}{15} x^3 + \frac{103}{140} x^4 + \ldots \right), \\
&\text{set $x=1-\delta_2/\delta_1$, valid for $\delta_1>\delta_2$, or} \nonumber \\
b &= \frac{\nu_2^2}{12 \delta_2^2} \left( 1 + x + \frac{13}{15} x^2 + \frac{11}{15} x^3 + \frac{103}{140} x^4 + \ldots \right),  \label{eqn:75} \\
&\text{set $x=1-\delta_1/\delta_2$, valid for $\delta_1<\delta_2$.} \nonumber
\end{align}

Expanding (\ref{eqn:72}) about $\zeta$ and keeping only first powers of $\zeta$, except for the term $b\zeta^2$ (because of the size of factor $b$) and setting
\begin{align}	\label{eqn:76}
\frac{d\delta_1}{d\zeta} = \delta \zeta, 
\end{align}
so (\ref{eqn:72}) becomes:
\begin{align}	\label{eqn:77}
B_{\rm Max} = F\cdot \frac{L_{21}}{L_{11}} \frac{\nu_2}{8} q \sqrt{1 - \zeta \left( 1 - \frac{2 \delta}{q} \frac{\partial q}{\partial \delta_1} \right) - b \zeta^2}. 
\end{align} 

\begin{center}	
\rule[0.3em]{0.3 \columnwidth}{0.5pt}
{\bf Page 529} 
\rule[0.3em]{0.3 \columnwidth}{0.5pt}
\end{center}
 
Naming $V^2$ the square of the maximum potential difference $2B_\text{Max}$ between the ends of the secondary circuit, then following (\ref{eqn:77}) $V^2$ takes the form of formula (\ref{eqn:64}), where there is 
\begin{align}	\label{eqn:78}
a = - \left( 1 - \frac{2 \delta}{q} \frac{\partial q}{\partial \delta_1} \right) = -1 + \frac{2\delta}{\delta_1-\delta_2}\left( \frac{\delta_2 \, {\rm log} \,\delta_1/\delta_2}{\delta_1-\delta_2} -1\right), 
\end{align} 
while $b$ has the value concluded from (\ref{eqn:73}),(\ref{eqn:74}) or (\ref{eqn:75}). 
Following (\ref{eqn:69}), one thus gains from the resonance curve measured quantitatively close to resonance: 
\begin{align}	\nonumber
\frac{V_m^2-V^2}{V_m^2}=b\eta^2
\end{align} 
Informative about the value of $b$, whereas the resonance frequency is
\begin{align}	\label{eqn:79}
\nu_1 = \nu_2 (1+\zeta_m) = \nu_2 \left( 1 - \frac{1 + \frac{2\delta}{\delta_1-\delta_2}\left(1-\frac{\delta_2}{\delta_1-\delta_2}{\rm log}\frac{\delta_1}{\delta_2}\right)}{2b} \right). 
\end{align} 

Whether the value of $\zeta_m$ is positive or negative, i.e.~if the resonance frequency is larger or smaller than the frequency $\nu_2$ of the secondary circuit, is impossible to tell without knowledge of the value $\delta$, i.e.~the dependence of $\delta_1$ on $\nu_1$. Anyway, $\delta$ has the order of magnitude of $\delta_1$ or $\delta_2$, such that $\zeta_m$, following (\ref{eqn:79}) and (\ref{eqn:73}), is of the order of magnitude
\begin{align}	\nonumber
\frac{1}{2} \left( \frac{\delta_1-\delta_2}{\nu_2} \right)^2 = \frac{1}{2} \left( \frac{\gamma_1-\gamma_2}{2\pi} \right)^2 ,
\end{align}
where $\gamma_1, \gamma_2$ are the logarithmic decrements of the two oscillation circuits (compare formula (\ref{eqn:30p}) p.~520). Even if it was $\gamma_1=1$, which corresponds to a large damping which would lead to only a weakly distinct resonance and it was $\gamma_2=0$, the resonance frequency of the primary circuit would only deviate by about $1-2$ percent from the frequency of the secondary circuit.\footnote{At my experiments (Ann.~d.~Phys. {\bf 9.}~p.~293. 1902), where I investigated the eigenoscillations of coils through resonance, the resonance was very strongly developed, such that the error was below \sfrac{1}{4} percent for sure.} 

\textit{b) $L_{11}$ in the primary circuit is changed.} Following (\ref{eqn:22}) p.~519, $p_{21}$ is constant. (\ref{eqn:57}) yields for the value of $t$ determined in (\ref{eqn:59}) and following (\ref{eqn:63})
\begin{align}	\label{eqn:80}
B_{\rm Max} = p_{21} F \frac{\nu_2}{4} \left(1+\frac{3}{2} \zeta\right) \left( q + \frac{\partial q}{\partial \delta_1}\cdot \frac{d\delta_1}{d\zeta}\right) \sqrt{1-b\zeta^2}, 
\end{align}

\begin{center}	
\rule[0.3em]{0.3 \columnwidth}{0.5pt}
{\bf Page 530} 
\rule[0.3em]{0.3 \columnwidth}{0.5pt}
\end{center}

\noindent
where $b$ has the same value (\ref{eqn:73}) as in case (a). In contrast,
\begin{align}	\label{eqn:81}
\frac{d\delta_1}{d\zeta} = \delta'\zeta
\end{align}
is here generally assumed to be different from the value of $d\delta_1/d\zeta$ in case a) (compare to the formula (\ref{eqn:76}) there). When expanding (\ref{eqn:80}) about the powers of $\zeta$ the form (\ref{eqn:64}) emerges, whereas now is
\begin{align}	\label{eqn:82}
a = +3 + \frac{2\delta'}{\delta_1-\delta_2}\left( \frac{\delta_2}{\delta_1-\delta_2}{\rm log} \frac{\delta_1}{\delta_2} - 1\right). 
\end{align}  

From the resonance curve one then again receives information about the dimension $b$, but now the resonance frequency following (\ref{eqn:65}) is
\begin{align}	\label{eqn:83}
\nu_1 = \nu_2(1+\zeta_m) = \nu_2 \left( 1 + \frac{3 - \frac{2\delta'}{\delta_1-\delta_2}\left(1-\frac{\delta_2}{\delta_1-\delta_2}{\rm log}\frac{\delta_1}{\delta_2}\right)}{2b} \right). 
\end{align}

When comparing this formula with the corresponding formula (\ref{eqn:79}) of case a) one can see that, when no big difference between the change coefficients $\delta$ and $\delta'$ is influencing the result, the resonance frequency in this case b) is greater than in case a).

\textit{2. The integral effect.}\footnote{It is also possible to determine the damping in the primary circuit from the integral effect on a metallic closed secondary circuit (which thus does not have eigenoscillations) (compare R.~Lindemann, Ann.~d.~Phys.~{\bf 12.}~p.~1012. 1903). This method is especially convenient, if one wants to study the change of the damping $\gamma_1$ with variation of $C_1$or $F$. I intend to go into more detail on another occasion and to publish the measurements.}
When observing the Joule heat generated by the secondary current $i_2$ in the middle of the secondary circuit ($i_2=i_2^0$), which can conveniently be done by a thermo element located there, it depends on the integral effect
\begin{align}	\nonumber
J = \int_0^\infty \! i_2^2 \, {d}t. 
\end{align}

Since from (\ref{eqn:3}), (\ref{eqn:8}) and (\ref{eqn:17}) it follows
\begin{align}	\nonumber
i_2^0=2C_2 \frac{ {d} V_2^h} {{d} t} \; \text{(compare with footnote on p.~519)},
\end{align}

\begin{center}	
\rule[0.3em]{0.3 \columnwidth}{0.5pt}
{\bf Page 531} 
\rule[0.3em]{0.3 \columnwidth}{0.5pt}
\end{center}

\noindent
equation (\ref{eqn:51}) on p.~523 yields:
\begin{align}	\nonumber
J &= \int_0^\infty \! 4C_2^2 B^2 (\nu_1+\nu_2)^2 \cos^2 \left( \frac{\nu_1+\nu_2}{2}t+\chi \right) \\
	&=2 C_2^2 (\nu_1+\nu_2)^2 \int_0^\infty B^2 (1+\cos[(\nu_1+\nu_2)t+2\chi]) \, {d}t. \nonumber 
\end{align}

One can neglect\footnote{See V.~Bjerknes, l.~c.~p.~137.} the effect of the second, with cos multiplied, part of the integral against the effect of the first part. Hence it becomes:
\begin{align}	\nonumber
J=2C_2^2(\nu_1+\nu_2)^2\int_0^\infty B^2 {d}t.
\end{align}

Here, when setting the value of $B$ following (\ref{eqn:55}), it follows, since
\begin{align}	\nonumber
\int e^{-(\delta_1+\delta_2)t} \cos&({\alpha t +\beta}) {d}t \\
	&= e^{-(\delta_1+\delta_2)t} \frac{\alpha \sin(\alpha t+\beta)-(\delta_1+\delta_2)\cos({\alpha t+\beta})}{\alpha^2+(\delta_1+\delta_2)^2} ,\nonumber\\
J=p_{21}^2 F^2 \frac{\nu_1^4}{2} &C_2^2 \frac{1}{(\nu_1-\nu_2)^2+(\delta_1-\delta_2)^2} \nonumber\\
	&\left \{ \frac{1}{2\delta_1}+\frac{1}{2\delta_2}-2 \frac{(\delta_1+\delta_2) \cos\beta-\alpha \sin\beta}{\alpha^2+(\delta_1+\delta_2)^2} \right \} . \nonumber 
\end{align}

Here, it is temporarily set following (\ref{eqn:56}):
\begin{align}	\nonumber
\begin{aligned} 
\alpha&=\nu_1 - \nu_2 , & \beta&=\frac{\delta_1 \nu_2+\delta_2 \nu_1}{\nu_1+\nu_2} \left ( \frac{1}{\nu_1}-\frac{1}{\nu_2} \right ). 
\end{aligned} 
\end{align}
Since it is now possible to neglect $\delta^2$ against $\nu^2$ it follows:
\begin{align}
J=p_{21}^2 F^2 \frac{\nu_1^4}{4} C_2^2 &\frac{1}{(\nu_1-\nu_2)^2+(\delta_1-\delta_2)^2} \nonumber \\
	&\left \{ \frac{1}{\delta_1} + \frac{1}{\delta_2} - 4 \frac{\delta_1+\delta_2+(\nu_1-\nu_2)^2 \frac{\delta_1\nu_2 + \delta_2 \nu_1}{(\nu_1+\nu_2)\nu_1 \nu_2}}{(\nu_1-\nu_2)^2+(\delta_1+\delta_2)^2} \right \} . \nonumber
\end{align}

Now, $(\nu_1-\nu_2)^2$ is to neglect against $(\nu_1+\nu_2)\nu_1 \nu_2 (\delta_1 + \delta_2) : \delta_1 \nu_2 + \delta_2 \nu_1$, but is not to neglect against $(\delta_1 + \delta_2)^2$ or against $(\delta_1-\delta_2)^2$. Hence, the last equation has to be written as:

\begin{center}	
\rule[0.3em]{0.3 \columnwidth}{0.5pt}
{\bf Page 532} 
\rule[0.3em]{0.3 \columnwidth}{0.5pt}
\end{center}

\begin{align}	\nonumber
J= p_{21}^2 F^2 \frac{\nu_1^4}{4} C_2^2 &\frac{1}{(\nu_1-\nu_2)^2+(\delta_1-\delta_2)^2} \nonumber \\
	&\left \{ \frac{1}{\delta_1} + \frac{1}{\delta_2} - \frac{4(\delta_1+\delta_2)}{(\nu_1-\nu_2)^2+(\delta_1+\delta_2)^2} \right \} . \nonumber
\end{align}
or: 
\begin{align}	\label{eqn:84}
J= p_{21}^2 F^2 \frac{\nu_1^4}{4} C_2^2 \frac{\delta_1+\delta_2}{\delta_1\delta_2}\cdot\frac{1}{(\nu_1-\nu_2)^2+(\delta_1+\delta_2)^2}. 
\end{align}

From this, it can be seen \textit{that by looking at the resonance curve of the integral effect one can obtain information about $\delta_1+\delta_2$}, which is a result which already Bjerknes\footnote{V.~Bjerknes, l.~c.~p.~148.} concluded. For more precise discussion about the way $\delta_1 + \delta_2$ is gained from the resonance curve one has to differentiate again between the two cases, either when varying $\nu_1$ then only $C_1$ is changed or only $L_{11}$.

\textit{a) The capacity $C_1$ in the primary circuit is changed.}
Following (\ref{eqn:62}) p.~525 in this case one has to set:
\begin{align}	\nonumber
p_{21}=\frac{1}{2} \frac{L_{21}}{L_{11}} \frac{\nu_2^2}{\nu_1^2}. 
\end{align}
Hence, (\ref{eqn:84}) becomes with respect to $\nu_2^2=1:L_{22} C_3$ [following (\ref{eqn:22})]:
\begin{align}	\nonumber
J=\frac{F^2}{16} \frac{L_{21}^2}{L_{11}^2L_{22}^2} \cdot \frac{\delta_1+\delta_2}{\delta_1\delta_2}\cdot\frac{1}{(\nu_1-\nu_2)^2+(\delta_1+\delta_2)^2}. 
\end{align}

Setting now $\nu_1=\nu_2(1+\zeta)$ following (\ref{eqn:63}) and $\mathrm{d} \delta_1 : \mathrm{d} \zeta = \delta \vartheta$ following (\ref{eqn:76}), it becomes
\begin{align}	\nonumber
J=\frac{F^2}{16} \frac{L_{21}^2}{L_{11}^2L_{22}^2} \cdot \frac{1}{\delta_1\delta_2(\delta_1+\delta_2)}\cdot \left( 1 - \frac{\delta_2\delta}{\delta_1(\delta_1+\delta_2)}\zeta - \left( \frac{\nu_2}{\delta_1+\delta_2}\right)^2\zeta^2\right), 
\end{align}
i.e., one gains again the form (\ref{eqn:64}) with the meaning of the coefficients
\begin{align}	\label{eqn:85}
\begin{aligned} 
a &= - \frac{\delta_2\delta}{\delta_1(\delta_1+\delta_2)}, & b&= \left(\frac{\nu_2}{\delta_1+\delta_2}\right)^2 = \left(\frac{2 \pi}{\gamma_1+\gamma_2}\right)^2. 
\end{aligned} 
\end{align}

Therefore, from the slope of the resonance curve it yields $\gamma_1+\gamma_2$, while the resonance frequency from (\ref{eqn:65}) follows to be
\begin{align}	\label{eqn:86}
\nu_1=\nu_2(1+\zeta_m)=\nu_2\left( 1 - \frac{\delta_2\delta(\delta_1+\delta_2)}{2\delta_1\nu_2^2}\right). 
\end{align}

Comparing the values of $b$ following (\ref{eqn:74}) or (\ref{eqn:75}) with the determined values of $b$ here following (\ref{eqn:85}) one can see that the latter is always slightly larger than the former, the most for 

\begin{center}	
\rule[0.3em]{0.3 \columnwidth}{0.5pt}
{\bf Page 533} 
\rule[0.3em]{0.3 \columnwidth}{0.5pt}
\end{center}

\noindent
the case $\delta_1=\delta_2$. Since now following p.~527 a bigger $b$ yields to a steeper (more distinct) resonance curve, \textit{the investigation of the integral effect yields to a sharper resonance then the one of the maximal effect,} a result, which was also already concluded by Bjerknes l.~c.~p.~165.\footnote{Also for $\nu_1=\nu_2$ the integral effect is much more dependent on the damping, since following (\ref{eqn:84}) it is inversely proportional to the power of $\delta_{1,2}$, than on the maximal amplitude, since following comment 1 p.~525 $V_\text{Max.}^2$ is proportional to $1/\delta_1^2$
.}

\textit{With the combination of the investigations of the resonance curve of the integral effect and the maximal effect one can calculate the logarithmic decrements of both oscillating circuits $\gamma_1, \gamma_2$ one by one,} since following (\ref{eqn:73}) and (\ref{eqn:85}) one calculates two different expressions composed of $\gamma_1$ and $\gamma_2$. This is shown by an experimentally accessible method to determine $\gamma_1$ and $\gamma_2$. 

If one calculated $\gamma_1$ and $\gamma_2$ or $\delta_1$ and $\delta_2$ using this method, \textit{then the relation between the resonance frequencies of the two effects (\ref{eqn:79}) and (\ref{eqn:86}) yields to a formula to calculate the changing coefficient $\delta$, i.e.~also for the exact value of the frequency $\nu_2$ of the secondary circuit.}

\textit{b) $L_{11}$ in the primary circuit is changed.} Following (\ref{eqn:22}) p.~519 $p_{21}$ is constant. Following (\ref{eqn:84}) it gets
\begin{align}	\nonumber
J \sim 1+ \left(4 - \frac{\delta_2 \delta'}{\delta_1(\delta_1+\delta_2)}\right)\zeta - \left( \frac{\nu_2}{\delta_1+\delta_2}\right)^2\zeta^2, 
\end{align}
where $\delta$ is the changing coefficient of $\delta_1$ defined in (\ref{eqn:81}). Therefore, from the resonance curve it yields $\nu_2:\delta_1+\delta_2$, the resonance curve is:
\begin{align}	\label{eqn:87}
\nu_1=\nu_2(1+\zeta_m)=\nu_2\left( 1 + \frac{4 - \frac{\delta_2\delta'}{\delta_1(\delta_1+\delta_2)}}{2 \nu_2^2:(\delta_1+\delta_2)^2}\right). 
\end{align}
This resonance frequency is in general higher than the resonance frequency in case (a) following (\ref{eqn:86}). A comparison with (\ref{eqn:82}) shows, that also in case (b) one can, through combination of the observation of the integral and the maximal effect, determine $\delta_1, \delta_2, \delta', \nu_2$ experimentally.

\begin{center}	
\rule[0.3em]{0.3 \columnwidth}{0.5pt}
{\bf Page 534} 
\rule[0.3em]{0.3 \columnwidth}{0.5pt}
\end{center}

\begin{center}
\textbf{IV. The Magnetic Coupling is Not Very Small.}\addcontentsline{toc}{subsubsection}{IV.~The Magnetic Coupling is Not Very Small}
\end{center}

When the magnetic coupling $k^2$ (formula (\ref{eqn:33})) can not be neglected against 1, we use the equation form (II), (\ref{eqn:25}) and (\ref{eqn:29}). For
\begin{align}	\label{eqn:88}
y = z - \frac{\vartheta_1+\vartheta_2}{2} 
\end{align}
(\ref{eqn:29}) yields:
\begin{align}	\label{eqn:89}
\left\{ \begin{aligned}
z^4 + z^2 \bigg[ \tau_1^2 &+ \tau_2^2-\frac{1}{2}(\vartheta_1 - \vartheta_2)^2\bigg] - z (\tau_1^2-\tau_2^2)(\vartheta_1-\vartheta_2) \\
	&+ \left[ \left( \frac{\vartheta_1-\vartheta_2}{2} \right)^2 + \tau_1^2 \right] \left[ \left( \frac{\vartheta_1-\vartheta_2}{2} \right)^2 + \tau_2^2 \right] \\ 
	&- k^2(\tau_1^2 + \vartheta_1^2)(\tau_2^2+\vartheta_2^2) = 0. 
\end{aligned} \right.
\end{align}

The four roots $z$ of this equation must have the form 
\begin{align}	\label{eqn:90}
\begin{aligned}
z_1 &= \beta + i \tau, & z_2 &= \beta - i \tau, & z_3 &= -\beta + i \tau', & z_4 &= - \beta + i \tau'. 
\end{aligned}
\end{align}
Since now (\ref{eqn:89}) is identical with
\begin{align}	\nonumber
(z-z_1)(z-z_2)(z-z_3)(z-z_4)=0,
\end{align}
the comparison of the coefficients of this last equation with the coefficients of the equation (\ref{eqn:89}) and using form (\ref{eqn:90}) yields:
\begin{align}	\label{eqn:91}
\left\{ \begin{aligned}
\tau^2 + \tau'^2 - 2 \beta^2 &= \tau_1^2 + \tau_2^2 - \frac{1}{2} (\vartheta_1-\vartheta_2)^2, \\
2 \beta (\tau^2 - \tau'^2) &= - (\vartheta_1-\vartheta_2)(\tau_1^2-\tau_2^2) = Q, \\
(\tau^2 + \beta^2)(\tau'^2 + \beta^2) &= \left(\frac{\vartheta_1-\vartheta_2}{2}\right)^4 + (\tau_1^2 + \tau_2^2) \left(\frac{\vartheta_1-\vartheta_2}{2}\right) \\
	&+ \tau_1^2\tau_2^2-k^2(\tau_1^2+\vartheta_1^2)(\tau_2^2+\vartheta_2^2). 
\end{aligned} \right.
\end{align}
When squaring the first of these equations and subtracting the last equation multiplied by 4, it becomes:
\begin{align}	\label{eqn:92}
\left\{ \begin{aligned}
(\tau^2-\tau'^2)^2-8\beta^2(\tau^2+\tau'^2) &= (\tau_1^2 - \tau_2^2)^2-2(\tau_1^2+\tau_2^2)(\vartheta_1-\vartheta_2)^2 \\
	&+ 4 k^2 (\tau_1^2+\vartheta_1^2)(\tau_2^2 + \vartheta_2^2) = P. 
\end{aligned} \right.
\end{align}

Now, there is a crucial difference, if this value $P$ is positive or negative. Anyway, with not too small magnetic coupling $k^2$, $P$ is positive. When we investigate close to the resonance $\tau_1=\tau_2$, it follows from the second of the equations (\ref{eqn:91}), that, anyway, one of the values $\beta$ or $\tau^2-\tau'^2$ has to be very small. Therefore, if $P$ is 

\begin{center}	
\rule[0.3em]{0.3 \columnwidth}{0.5pt}
{\bf Page 535} 
\rule[0.3em]{0.3 \columnwidth}{0.5pt}
\end{center}

\noindent
positive, following (\ref{eqn:92}), it has to be the value $\beta$ itself and we receive for $\tau_1=\tau_2$:
\begin{align}	\label{eqn:93}
\begin{aligned}
\beta &= 0, & \tau^2-\tau'^2&=\sqrt{P} & &\text{for} &\tau_1 &= \tau_2. 
\end{aligned}
\end{align}

The four roots $y$ have therefore common real parts, but the imaginary parts are different. \textit{Therefore, with not too small magnetic coupling in case of resonance $\tau_1=\tau_2$ it creates two oscillations with differing periods and differing\footnote{M. Wien (Wied.~Ann.~{\bf 61.}~p.~177.~1897; also compare Ann.~d.~Phys.~{\bf 8.}~p.~695.~1902) had derived, that the case of magnetic coupling is not essentially different to the case of electric coupling. Therefore, the damping of the two waves are equal. Although, this result is only valid, if the coupling $k^2$ (following Wien's labeling $\varrho_1\varrho_2$) is small against 1, since otherwise Wien's assumptions are not allowed. This here concluded result is in accordance with A.~Oberbeck (Wied.~Ann.~{\bf 55.}~p.~631.~1895).} damping,} since it is, when neglecting $\vartheta^2$ against $\tau^2$, i.e.\footnote{Compare above p.~521, equation (\ref{eqn:32}).} the square of the logarithmic decrement towards $4\pi^2$, which is always allowed if the resonance becomes distinctly evident at all:
\begin{align}	\nonumber
\frac{1}{y_1} &= \frac{1}{-\frac{\vartheta_1+\vartheta_2}{2}+i \tau}=-\frac{\vartheta_1+\vartheta_2}{2 \tau^2}-\frac{i}{\tau} , \\
\nonumber
\frac{1}{y_2} &= \frac{1}{-\frac{\vartheta_1+\vartheta_2}{2}+i \tau'}=-\frac{\vartheta_1+\vartheta_2}{2 \tau'^2}-\frac{i}{\tau'}.
\end{align}

\textit{Hence, the slower oscillation} (period $\sim \tau$) \textit{has a smaller absolute damping and a smaller logarithmic decrement than the faster oscillation} (period $\sim \tau', \tau'<\tau$). 

If contrariwise $P$ is negative, which can occur with a very small coupling $k^2$ close to the resonance, $\tau^2-\tau'^2$ would become small, $\beta$ large. With neglecting $k^2$ (\ref{eqn:91}) would then yield: $\tau=\tau_1, \tau'=\tau_2, \beta=-\frac{1}{2} (\vartheta_1-\vartheta_2)$, i.e.~following (\ref{eqn:88}) and (\ref{eqn:90}) it would be
\begin{align}	\nonumber
\begin{aligned} 
y_1&=-\vartheta_1+i\tau_1, & y_2&=-\vartheta_1-i\tau_1, \\
\nonumber
y_3&=-\vartheta_2+i\tau_2, & y_4&=-\vartheta_1-i\tau_2.
\end{aligned} 
\end{align}

Therefore, this yields the case already treated and solved at II.

\begin{center}	
\rule[0.3em]{0.3 \columnwidth}{0.5pt}
{\bf Page 536} 
\rule[0.3em]{0.3 \columnwidth}{0.5pt}
\end{center}

\textit{We therefore have to assume $P$ to be positive.} The general solution of (\ref{eqn:91}), (\ref{eqn:92}) proves to be simple close to the resonance $\tau_1=\tau_2$, since $Q$ there is small, i.e.~also $\beta^2$ is small against $\tau^2-\tau'^2$. Hence, in (\ref{eqn:92}) one can substitute in the factor of $8\beta^2$ for $\tau^2-\tau'^2$ the according approximate value $\tau^2+\tau'^2=\tau_1^2+\tau_2^2$, which is received for $\beta=0$ from the first equation of (\ref{eqn:91}) and with neglecting $\frac{1}{2}(\vartheta_1-\vartheta_2)^2$ against $\tau_1^2+\tau_2^2$. Then (\ref{eqn:92}) yields:
\begin{align}	\label{eqn:94}
(\tau^2-\tau'^2)^2-8\beta^2(\tau_1^2+\tau_2^2) = P. 
\end{align}

The second of the equations (\ref{eqn:91}) and (\ref{eqn:94}) one can now readily solve for the two unknown $\tau^2-\tau'^2$ and $\beta^2$. One gets:
\begin{align}	\label{eqn:95}	
\left\{ \begin{aligned}
(\tau^2-\tau'^2)^2 &= \frac{P+\sqrt{P^2+8 Q^2(\tau^2_1+\tau^2_2)}}{2}, \\ 
8 \beta^2(\tau^2_1+\tau^2_2) &=\frac{-P+\sqrt{P^2+8Q^2(\tau^2_1+\tau^2_2)}}{2}. 
\end{aligned} \right.
\end{align}
	
Since close to the resonance $8 Q^2(\tau^2_1+\tau^2_2)$ is small against $P^2$, for (\ref{eqn:95}) one can set:
\begin{align}	\label{eqn:96}
\left\{ \begin{aligned}
\tau^2-\tau'^2 &= \sqrt{P}\left( 1 + \frac{Q^2 (\tau_1^2 + \tau_2^2)}{P^2}\right), \\
\beta &= - \frac{Q}{2\sqrt{P}}, 
\end{aligned} \right.
\end{align}
while the first one of the equations (\ref{eqn:91}) yields:
\begin{align}	\label{eqn:97}
\tau^2 + \tau'^2 = \tau_1^2+\tau_2^2 + \frac{Q^2}{2P}. 
\end{align}

Further, following (\ref{eqn:25}), (\ref{eqn:88}) and (\ref{eqn:90}) it is valid (for arbitrary values of the coupling $k^2$): 
\begin{align}	\label{eqn:98}
\left\{ \begin{aligned}
V_1 &= A_1 e^{t/y_1} + A_2 e^{t/y_2} + A_3 e^{t/y_3} + A_4 e^{t/y_4}, \\ 
V_2 &= B_1 e^{t/y_1} + B_2 e^{t/y_2} + B_3 e^{t/y_3} + B_4 e^{t/y_4}. 
\end{aligned} \right.
\end{align}
\begin{align}	\label{eqn:99}
\left\{ \begin{aligned}
y_1 &= \beta - \frac{\vartheta_1+\vartheta_2}{2} + i \tau, & y_2 &= \beta - \frac{\vartheta_1+\vartheta_2}{2} - i \tau, \\
y_3 &= -\beta - \frac{\vartheta_1+\vartheta_2}{2} + i \tau', & y_4 &= -\beta - \frac{\vartheta_1+\vartheta_2}{2} - i \tau'. 
\end{aligned} \right.
\end{align}

\begin{center}	
\rule[0.3em]{0.3 \columnwidth}{0.5pt}
{\bf Page 537} 
\rule[0.3em]{0.3 \columnwidth}{0.5pt}
\end{center}

From (\ref{eqn:1}) it follows:
\begin{align}	\label{eqn:100}
-\frac{i_1}{C_1} = \frac{A_1}{y_1} e^{t/y_1} + \frac{A_2}{y_2} e^{t/y_2} + \frac{A_3}{y_3} e^{t/y_3} + \frac{A_3}{y_3} e^{t/y_3}, 
\end{align}
from (\ref{eqn:3}) and (\ref{eqn:8}):
\begin{align}	\label{eqn:101}
\pi \frac{i_2}{\mathfrak{C}_2 h} = \frac{B_1}{y_1} e^{t/y_1} + \frac{B_2}{y_2} e^{t/y_2} + \frac{B_3}{y_3} e^{t/y_3} + \frac{B_3}{y_3} e^{t/y_3}. 
\end{align}

Hence, the initial conditions (\ref{eqn:38}) yield:
\begin{gather*} \label{eqn:102}
\begin{align}	
\begin{aligned} 
\sum A_n &= F, & \sum \frac{A_n}{y_n} &= 0, & \sum B_n &= 0, & \sum \frac{B_n}{y_n} &= 0, 
\end{aligned} 
\end{align} \\
n = 1, 2, 3, 4. 
\end{gather*}

Now, following the second\footnote{From this it also follows for the case $\tau_1=\tau_2$ and $\vartheta_1=\vartheta_2$ that it is $A_1:B_1=\tau^2_2-\tau^2:p_{21}\tau_2^2$, thus negative, since $\tau>\tau_2$, and that it is $A_3:B_3=\tau_2^2-\tau'^2:p_{21}\tau_2^2$, i.e.~positive, since $\tau'<\tau_2$. With respect to (\ref{eqn:100}) and (\ref{eqn:101}) this yields that the oscillation $\tau$ is rectified in $i_1$ and $i_2$, but the oscillation $\tau'$ runs reverse in $i_1$ and $i_2$. Hence, one can possibly measure the period $\tau$ with a resonance line set up close to $i_1$ easier (comp.~P.~Drude, Ann.~d.~Phys.~{\bf 9.}~p.~611.~1902) than the period $\tau'$.}
of the equations (\ref{eqn:27}):
\begin{align}	\nonumber
B_n \left( \frac{\tau_2^2+\vartheta^2_2}{y_n}+2\vartheta_2+y_n \right)=p_{21}(\tau_2^2+\vartheta_2^2)\frac{A_n}{y_n}.
\end{align}

With addition of these four equations constructed from $n=1,2,3,4$ and due to the equation (\ref{eqn:102}) it results:
\begin{align}	\label{eqn:103}
\sum B_n y_n = 0.
\end{align}

Furthermore, it follows from (\ref{eqn:27}):
\begin{align}	\nonumber
B_n(\tau_2^2+\vartheta_2^2+2\vartheta_2y_n+y_n^2)=p_{21}(\tau_2^2+\vartheta_2^2) A_n.
\end{align}

With addition of these four equations it follows due to (\ref{eqn:102}) and (\ref{eqn:103}):
\begin{align}	\label{eqn:104}
\sum B_n y_n^2 = p_{21} (\tau_2^2 + \vartheta_2^2) F. 
\end{align}

From (\ref{eqn:102}), (\ref{eqn:103}) and (\ref{eqn:104}) is is immediately possible to describe $B$ as quotients of the determinant, e.g.~it is
\begin{align}	\nonumber
\frac{B_1}{y_1} \cdot 
\begin{vmatrix}
1, & 1, & 1, & 1 \\
y_1, & y_2, & y_3, & y_4 \\
y_1^2, & y_2^2, & y_3^2, & y_4^2 \\
y_1^3, & y_2^3, & y_3^3, & y_4^3 
\end{vmatrix} 
= -p_{21} (\tau_2^2 + \vartheta_2^2) F \cdot 
\begin{vmatrix}
1, & 1, & 1 \\
y_2, & y_3, & y_4 \\
y_2^2, & y_3^2, & y_4^2
\end{vmatrix}. 
\end{align}

\begin{center}	
\rule[0.3em]{0.3 \columnwidth}{0.5pt}
{\bf Page 538} 
\rule[0.3em]{0.3 \columnwidth}{0.5pt}
\end{center}

Following (\ref{eqn:45}) this yields
\begin{align}	\nonumber
B_1(y_1-y_2)(y_1-y_3)(y_1-y_4)=+p_{21}(\tau_2^2+\vartheta_2^2)Fy_1,
\end{align}
and analogously: 
\begin{align}	\nonumber
B_2(y_2-y_3)(y_2-y_4)(y_2-y_1)=p_{21}(\tau_2^2+\vartheta_2^2)Fy_2, \\
\nonumber
B_3(y_3-y_4)(y_3-y_1)(y_3-y_2)=p_{21}(\tau_2^2+\vartheta_2^2)Fy_3, \\
\nonumber
B_4(y_4-y_1)(y_4-y_2)(y_4-y_3)=p_{21}(\tau_2^2+\vartheta_2^2)Fy_4.
\end{align}

Hence, following (\ref{eqn:99}) it is
\begin{align}	\nonumber
B_1=p_{21}(\tau_2^2+\vartheta_2^2)F\frac{\beta-\frac{\vartheta_1+\vartheta_2}{2}+i\tau}{2i\tau(2\beta+i(\tau-\tau'))(2\beta+i(\tau+\tau'))}.
\end{align}

Now $(2\beta)^2$ is always to neglect against $(\tau+\tau')$. Hence it is
\begin{align}	\nonumber
\frac{1}{2\beta+i(\tau+\tau')}=\frac{1+i\frac{2\beta}{\tau+\tau'}}{i(\tau+\tau')},
\end{align}
if one neglects $\vartheta^2_2$ against $\tau_2^2$ as well as $\beta^2$ against $\tau^2$, which is allowed, if it is a resonance at all, and one uses the abbreviation: 
\begin{align}	\label{eqn:105}
\delta = \frac{\vartheta_1+\vartheta_2}{2} + \beta \frac{\tau-\tau'}{\tau + \tau'}. 
\end{align}
\begin{align}	\label{align}
\left\{ \begin{aligned}
B_1 &= -p_{21} \tau_2^2 F \frac{\tau - \tau' - 2 \frac{\beta}{\tau}\delta + i \left( 2 \beta + \frac{\tau - \tau'}{\tau} \delta \right)}{2 (\tau + \tau') [4 \beta^2 + (\tau - \tau')^2]}, \\
B_2 &= -p_{21} \tau_2^2 F \frac{\tau - \tau' - 2 \frac{\beta}{\tau}\delta - i \left( 2 \beta + \frac{\tau - \tau'}{\tau} \delta \right)}{2 (\tau + \tau') [4 \beta^2 + (\tau - \tau')^2]}, \\
B_3 &= +p_{21} \tau_2^2 F \frac{\tau - \tau' - 2 \frac{\beta}{\tau}\delta + i \left( 2 \beta + \frac{\tau - \tau'}{\tau} \delta \right)}{2 (\tau + \tau') [4 \beta^2 + (\tau - \tau')^2]}, \\
B_4 &= +p_{21} \tau_2^2 F \frac{\tau - \tau' - 2 \frac{\beta}{\tau}\delta - i \left( 2 \beta + \frac{\tau - \tau'}{\tau} \delta \right)}{2 (\tau + \tau') [4 \beta^2 + (\tau - \tau')^2]}. 
\end{aligned} \right.
\end{align}

Here, $\beta^2$ is not neglected against $(\tau-\tau')\tau$ in order to not exclude cases with small coupling at first, where $\tau-\tau'$ is small. 

Now with neglecting of $\beta^2$ against $\tau^2$ following (\ref{eqn:99}) it is:
\begin{align}	\nonumber
\frac{1}{y_1}=\frac{\beta}{\tau^2}-\frac{\vartheta_1+\vartheta_2}{2\tau^2}-\frac{i}{\tau},
\end{align}

\begin{center}	
\rule[0.3em]{0.3 \columnwidth}{0.5pt}
{\bf Page 539} 
\rule[0.3em]{0.3 \columnwidth}{0.5pt}
\end{center}

\noindent
hence following (\ref{eqn:98}) it becomes:
\begin{align}	\label{eqn:107}
B_2 = - \frac{p_{21} \tau_2^2 F \varPhi}{2 (\tau + \tau')[4 \beta^2 + (\tau - \tau')^2]}, 
\end{align}
while $\varPhi$ stands for:
\begin{align}	\label{eqn:108}
\left\{ \begin{aligned}
\varPhi &= e^{-\frac{t}{\tau^2} \left( \frac{\vartheta_1+\vartheta_2}{2}-\beta\right)}\cdot \left[ \left(\tau - \tau' -2 \frac{\beta \delta}{\tau}\right) \cos \frac{t}{\tau} + \left( 2 \beta + \frac{ \tau-\tau'}{\tau} \delta \right) \sin\frac{t}{\tau} \right] \\
	&- e^{-\frac{t}{\tau'^2} \left( \frac{\vartheta_1+\vartheta_2}{2}-\beta\right)}\cdot \left[ \left(\tau - \tau' -2 \frac{\beta \delta}{\tau'}\right) \cos \frac{t}{\tau'} + \left( 2 \beta + \frac{ \tau-\tau'}{\tau'} \delta \right) \sin\frac{t}{\tau'} \right]. 
\end{aligned} \right.
\end{align}

If we first ask, \textit{for which ratio $\tau_1:\tau_2$ the strongest induction will occur}, i.e.~$V_2$ becomes a maximum, it is to differentiate, the same way as above at p.~526, if with variation of $\tau_1$, either, the capacity $C_1$ is variated or the self induction $L_{11}$. Besides that, for estimation of the ratios one can read out the following from (\ref{eqn:107}): With variation of $\tau_1$ (while $\tau_2$ remains fixed) $V_2$ is changing mainly because of the denominator $4\beta^2+(\tau-\tau')^2$ and because of the numerator $\varPhi$, while $p_{21}$ is changing only negligible. Since $\beta$ is now very small close to the resonance and since $\delta$ is also small, $\varPhi$ has a value\footnote{A partition of $\varPhi$ in a product 
\begin{align}	\nonumber
\sin\frac{1}{2}t \left (\frac{1}{\tau}+\frac{1}{\tau'} \right ) \cdot \sin \frac{1}{2} t \left ( \frac{1}{\tau}-\frac{1}{\tau'} \right)
\end{align}
has only a physical meaning (beat), if $\tau$ is almost equal to $\tau'$. Here, we cannot assume this for stronger coupling.} which is constantly smaller than $2(\tau-\tau')$; therefore we can set:
\begin{align}	\label{eqn:109}
\begin{aligned} 
\varPhi &= 2 \varrho (\tau - \tau'), & \varrho &< 1.
\end{aligned} 
\end{align}

The value of the real fraction $\varrho$ is not varying strongly when $\tau_1:\tau_2$ is changing. Therefore, with neglecting of $4\beta^2$ against $(\tau-\tau')^2$, which is allowed with strong coupling, (\ref{eqn:107}) becomes:
\begin{align}	\label{eqn:110}
V_2 = \frac{p_{21} \tau_2^2 F \varrho}{\tau^2 - \tau'^2}. 
\end{align}

Therefore, maximal induction occurs, if the periods of both induced oscillations get as close to each other as possible, 

\begin{center}	
\rule[0.3em]{0.3 \columnwidth}{0.5pt}
{\bf Page 540} 
\rule[0.3em]{0.3 \columnwidth}{0.5pt}
\end{center}

\noindent
i.e.~if $\tau^2-\tau'^2$ is a minimum. Following (\ref{eqn:92}) now it is approximately to set:
\begin{align}	\label{eqn:111}
\tau^2 -\tau'^2 = \sqrt{(\tau_1^2 - \tau_2^2)^2 + 4 k^2 \tau_1^2 \tau_2^2}, 
\end{align}
from where it follows that for $\tau_1=\tau_2$ the minimum of $\tau^2-\tau'^2$ is occurring. \textit{Therefore, the strongest induction occurs close to the isochronicity $\tau_1=\tau_2$.} Likewise, it follows from (\ref{eqn:110}) and (\ref{eqn:111}), \textit{that the resonance curve becomes flatter as the coupling becomes stronger,\footnote{This follows both from the measurements of F.~Kiebitz, Ann.~d.~Phys.~{\bf 5.}~p.~895.~1901, which interpreted the flatter resonance curve as a increase of the damping, and also from a number of my observations where I received the higher resonance the weaker I choose the coupling.}} since when $k^2$ is large $\tau^2-\tau^{'2}$ is changing proportionally less with change of $\tau_1^2-\tau_2^2$ than when $k^2$ is small. 

\textit{For the measurement of oscillation periods one therefore has to work with the smallest coupling possible.}

\begin{center}
\textbf{V.~The effectivity of Tesla coils.} 
\end{center}

Following the recent results we can assume the resonance case $\tau_1=\tau_2$ and we then obtain, following (\ref{eqn:92}) and (\ref{eqn:110}), since following (\ref{eqn:91}) $\beta$ has to be strictly equals zero, for the potential at the end of the Tesla coil the value:
\begin{align}	\nonumber
V_2=\frac{p_{21} F \varrho}{2\sqrt{k^2 - \left( \frac{\vartheta_1-\vartheta_2}{2}\right)^2}}
	=\frac{\varrho F}{4} \frac{L_{21} C_1}{L_{22} C_2}\frac{1}{\sqrt{k^2 - \left( \frac{\vartheta_1-\vartheta_2}{2}\right)^2}}. 
\end{align}

For $p_{21}$ the value following equation (\ref{eqn:22}) is inserted. 

Since now following (\ref{eqn:33}) it is 
\begin{align}	\nonumber
L_{21}=k\sqrt{\frac{L_{11}L_{22}L_{21}}{L_{12}}},
\end{align}
it follows
\begin{align}	\nonumber
V_2 = \frac{\varrho F}{4} \frac{C_1}{C_2}\frac{k}{\sqrt{k^2 - \left( \frac{\vartheta_1-\vartheta_2}{2}\right)^2}} \cdot \sqrt{\frac{L_{11}}{L_{22}} \cdot \frac{L_{21}}{L_{12}}}, 
\end{align}
or since in the resonance case it is $L_{11}C_1=L_{22}C_2$ and taking (\ref{eqn:32}) into account:\footnote{This formula is equal to the formula (\ref{eqn:20}) derived by A.~Oberbeck (Wied.~Ann.~{\bf 55.}~p.~629.~1895) except for the factor 
\begin{align}
\sqrt{\frac{L_{21}}{L_{12}}} \cdot \sqrt{\frac{k^2}{k^2-\left (\frac{\gamma_1-\gamma_2}{2\pi} \right )^2}}. \nonumber
\end{align}} 

\begin{center}	
\rule[0.3em]{0.3 \columnwidth}{0.5pt}
{\bf Page 541} 
\rule[0.3em]{0.3 \columnwidth}{0.5pt}
\end{center}

\begin{align}	\label{eqn:112}
V_2 = \frac{\varrho}{4} F \sqrt{\frac{C_1}{C_2} \frac{L_{21}}{L_{12}} \cdot \frac{k^2}{k^2- \left( \frac{\gamma_1-\gamma_2}{2 \pi} \right)^2}}. 
\end{align}

From (\ref{eqn:108}) the value for $\varrho$ is:
\begin{align}	\nonumber
2 \varrho &= e^{-\frac{t}{\tau^2} \cdot \frac{\vartheta_1+\vartheta_2}{2}} \left( \cos \frac{t}{\tau} + \frac{\delta}{\tau} \sin \frac{t}{\tau} \right) \\
	&- e^{-\frac{t}{\tau'^2} \cdot \frac{\vartheta_1+\vartheta_2}{2}} \left( \cos \frac{t}{\tau'} + \frac{\delta}{\tau'} \sin \frac{t}{\tau'} \right). \nonumber 
\end{align}

Following (\ref{eqn:105}) it is $\delta=\vartheta_1+\vartheta_2/2$, and one can set
\begin{align}	\nonumber
\cos\frac{t}{\tau}+\frac{\delta}{\tau}\sin\frac{t}{\tau}=\cos\frac{t-\delta}{\tau}, 
\end{align}
since $(\delta:\tau)^2$ can be neglected against 1. Hence, it becomes
\begin{align}	\nonumber
2 \varrho = e^{-\frac{t\delta}{\tau^2}} \cdot \cos \frac{t - \delta}{\tau} - e^{-\frac{t\delta}{\tau'^2}} \cdot \cos \frac{t - \delta}{\tau'}. 
\end{align}

When introducing a new time $t_1=t-\delta$, it becomes
\begin{align}	\nonumber
2 \varrho = e^{-\frac{\delta}{\tau}\left(\frac{t_1+\delta}{\tau}\right)} \cdot \cos \frac{t_1}{\tau} - e^{-\frac{\delta}{\tau'}\left(\frac{t_1+\delta}{\tau'}\right)} \cdot \cos \frac{t_1}{\tau'}.  
\end{align}
Since the maximum values of $\varrho$ are reached at the time $t_1$ which is about the same as $T_1$ or larger than $T_1$, one can neglect $\delta$ in comparison to $t_1$ in the exponential part and one gets, when then again writing $t$ instead of $t_1$:
\begin{align}	\label{eqn:113}
2 \varrho = e^{-\frac{t\delta}{\tau^2}} \cdot \cos \frac{t}{\tau} - e^{-\frac{t\delta}{\tau'^2}} \cdot \cos \frac{t}{\tau'}. 
\end{align}
In order to find the absolute maximum of $V_2$ one has to find that time $t$, for that $\varrho$ becomes a absolute maximum. Before we do that, some general notifications to (\ref{eqn:112}) are made. 

When we can neglect the damping $\gamma_1$ and $\gamma_2$, $\varrho_\text{Max}$ would have the value 1 and $V_2$ would be, following (\ref{eqn:112}) totally independent of the coupling, if it is possible to set $L_{12}=L_{21}$ like in linear circuits. Following

\begin{center}	
\rule[0.3em]{0.3 \columnwidth}{0.5pt}
{\bf Page 542} 
\rule[0.3em]{0.3 \columnwidth}{0.5pt}
\end{center}

\noindent 
p.~518, $L_{21}:L_{12}$ is growing with increasing coupling, though. \textit{Therefore, for that reason and due to existing damping the excitement of the Tesla coil has to increase with increasing coupling.} 

\textit{Not inductive (dead) self induction of the primary circuit is harmful for the best effectivity of the Tesla coil for two reasons.} First, the capacity $C_1$ of the Tesla coil becomes lower, since with the Tesla coil used here the oscillation period is given. Second, the coupling becomes smaller, which concludes from (\ref{eqn:33}). 

Since, following (\ref{eqn:112}) $V_2$ is proportional to $\sqrt{C_1:C_2}$, \textit{for best construction of a Tesla coil the primary circuit has to be made of one turn of thick filament} 
(since then $L_{11}$ is as small as possible and $C_1$ large as possible for a given $T_1$), \textit{while the secondary coil has to have a relatively small capacity in comparison to $C_1$.} Hence, only one open coil 
layer as secondary circuit is relatively advantageous due to resulting small $C_2$, at least better than a secondary coil existing out of a low number (2-10) of coil 
layers. This is because the small distance between the 
ends of the coil windings, which carry contrary polarity give a magnified capacity $C_2$. This result is indeed confirmed by the experiments I performed (measuring the potential through looking at the spark gap or qualitative through illumination of vacuum rods). 

One therefore has to use coils with many more 
turns to achieve similar favorably small values of capacity $C_2$ again. 

It is $h$ the height, $r$ the radius, $g$ the pitch, $\delta$ the thickness of the filament and $n$ the absolute number of coil windings of the Tesla coil. Then, the wavelength $\lambda$ of the eigenoscillation is determined by
 \begin{align}	\nonumber
 \sfrac{1}{2}\lambda=l\cdot f(h/r, g/\delta),
 \end{align}
where $l=2r\pi n$ is the length of the filament of the Tesla coil and $f$ is a factor depending on $h/r, g/\delta$ and the nature of the coil core, for which I noted tables earlier.\footnote{P.~Drude, Ann.~d.~Phys.~{\bf 9.}~p.~322.~1902.} 

When naming the radius of the circle which resembles the primary circuit $r_1$ (we assume a circle of one coil winding following the thoughts made above), 

\begin{center}	
\rule[0.3em]{0.3 \columnwidth}{0.5pt}
{\bf Page 543} 
\rule[0.3em]{0.3 \columnwidth}{0.5pt}
\end{center}

\noindent
and the radius of the filament $\varrho_1$ (half the filament's thickness), for the self induction of the primary circuit it thus is:\footnote{M.~Wien, Wied.~Ann.~{\bf 53.}~p.~931.~1894.} 
\begin{align}	\nonumber
L_{11}=4 \pi r_1 [(\text{lognat} \, 8 r_1/\varrho_1)-2],
\end{align}
therefore, it has to be for $C_1$, since it is
\begin{align}	\nonumber
2 \pi \sqrt{L_{11}C_1} = \lambda_1, 
\end{align}
 when $C_2$ is measured following the electrostatic measure: 
 \begin{align}	\nonumber
C_1 = \frac{\lambda_1^2}{4\pi^2L_{11}} = \frac{r^2n^2f^2}{\pi r_1 ({\rm log} \, 8 r_1/\varrho_1 - 2)}. 
 \end{align}
 The capacity $C_2$ of the Tesla coil (following the electrostatic measure) we can also write in the form of:
 \begin{align}	\nonumber
 C_2=r \varphi (h/r g/\delta). 
 \end{align}
Since now following (\ref{eqn:17}) and (\ref{eqn:18}) p.~517 $L_{21}:L_{12}$ and also $k^2$ following (\ref{eqn:33}) are not depending on $n$, but only on $h/r, r_1/r$, and since the maximum of $\varrho$  in time is only depending on $k^2, \gamma_1$ and $\gamma_2$, we gain from (\ref{eqn:112}) the sentence: 

\textit{With constant ratios $h/r, r_1/r, r_1/\varrho_1$ and with constant decrements $\gamma_1$ and $\gamma_2$ the maximum potential of the Tesla coil is proportional with its number of windings.} 

Now, it is to investigate the, especially for practice, important question: \textit{For which ratio ${h}/{r}$ of the Tesla coil is the effect the largest with given coil winding $n$?} I will give the answer to this question later in an experimental way. Approximately, the best ratio ${h}/{r}$ is between 1.5 and 3. 

If you neglect $(\gamma_1-\gamma_2)^2:4\pi^2$ to $k^2$, which is usually allowed with Tesla coils, (\ref{eqn:112}) can be written, with using the rearrangement
 \begin{align}	\nonumber
C_2 = r_1 \cdot r/r_1 \cdot \varphi(h/r,g/\delta) = L_{11} \cdot \chi (h/r,r/r_1,r_1/\varrho_1,g/\delta) \\ 
V_2 = \varrho F \sqrt{\frac{C_1}{L_{11}}} \cdot \psi(h/r,r/r_1,r_1/\varrho_1,g/\delta).  \label{eqn:114}
 \end{align}

Following (\ref{eqn:22}) and (\ref{eqn:30p}) now it is 
\begin{align}	\nonumber
\gamma_1 = \frac{\omega_1}{2} \sqrt{\frac{C_1}{L_{11}}}.
\end{align}
Therefore it becomes
\begin{align}	\tag{114'}\label{eqn:114p}
V_2 = 2 \varrho F \frac{\gamma_1}{w_1} \psi(h/r,r/r_1,r_1/\varrho_1,g/\delta). 
\end{align}

\begin{center}	
\rule[0.3em]{0.3 \columnwidth}{0.5pt}
{\bf Page 544} 
\rule[0.3em]{0.3 \columnwidth}{0.5pt}
\end{center}

$\omega_1$ is almost only consisting of the spark resistance.
This one is independent of $L_{11}$, but dependent on $F$ and $C_1$. It decreases with increasing $C_1$.\footnote{Compare R.~Lindemann. Ann.~d.~Phys.~{\bf 12.}~p.~1012.~1903.} $\varrho$ decreases with increasing $\gamma_1$. For \textit{weak} coupling one can view $\varrho \gamma_1$ approximately\footnote{Here $\gamma_2$ is considered small to $\gamma_1$.} as constant (compare section VI), whereas $ \varrho$ should be understood as the maximum reachable value over time. Therefore it follows from (\ref{eqn:114p}): \textit{The effectiveness of a not very strongly coupled Tesla coil is, with constant ratios ${h}/{r}, {r}/{r_1}, {r_1}/{\varrho_1}, {g}/{\delta}$, only dependent on the capacity $C_1$ of the primary circuit, no matter if the coil has small dimensions and many windings or large dimensions and less windings.}
In contrast, with very strong coupling $\delta \gamma_1$ increases with $\gamma_1$. \textit{Therefore, then a large $n$ and a small $r$ is slightly advantageous to a small $n$ and a large $r$.} 

\textit{In any case it depends on that $C_1$ necessary for the resonance is as large as possible.} 

\textit{The dependence of the Tesla effect of the primary spark potential F is small in between certain borders; } certain experiments revealed that to me and it also concludes from (\ref{eqn:114p}), since $\omega_1$ increases with the spark gap, therefore also with $F$. 

An embedding of the coil in petroleum would be disadvantageous due to the related raise of the capacity but could be advantageous due to the possibility of higher coupling $k^2$ ($r$ approximately equals $r_1$). The advantage will predominate the disadvantage when there is strong coupling; it will not when there is weak coupling though. 

Now it will be investigated, how the maximum value of the factor $\varrho$ in equation (\ref{eqn:112}) depends on the damping $\gamma_1, \gamma_2$ and the coupling $k^2$.

\begin{center}
\textbf{VI.~Dependence of the Tesla effect on damping and coupling.} 
\end{center}

Following (\ref{eqn:92}) and since $\beta=0$, following (\ref{eqn:91}), it is for the resonance case $\tau_1=\tau_2$:
\begin{align}	\nonumber
\tau^2-\tau'^2 = \sqrt{P} = 2 \tau_1^2 \sqrt{k^2-\left(\frac{\vartheta_1-\vartheta_2}{\tau_1}\right)^2} = 2 \tau_1^2 k', 
\end{align}
whereas $k'$ is a new abreviation for
\begin{align}	\label{eqn:115}
k'^2 = k^2 - \left(\frac{\vartheta_1-\vartheta_2}{\tau_1}\right)^2 = k^2 - \left(\frac{\gamma_1-\gamma_2}{2\pi}\right)^2. 
\end{align}

\begin{center}	
\rule[0.3em]{0.3 \columnwidth}{0.5pt}
{\bf Page 545} 
\rule[0.3em]{0.3 \columnwidth}{0.5pt}
\end{center}

\noindent 
Since $\frac{1}{2}(\vartheta_1-\vartheta_2)^2$ can be neglected against $2 \tau^2_1$, following (\ref{eqn:91}) it is:
\begin{align}	\nonumber
\tau^2+\tau'^2=2\tau^2_1.
\end{align}
From this it follows
\begin{align}	\label{eqn:116}
\begin{aligned} 
\tau^2 &= \tau_1^2 (1+k'), & \tau'^2 &= \tau_1^2(1-k'). 
\end{aligned} 
\end{align}
According to (\ref{eqn:115}) $k'^2$ can only be seen directly as coupling coefficient when there is not too small coupling $k^2$. In general, one can define $k'$ from $(\ref{eqn:116})$, namely
\begin{align}	\label{eqn:117}
\begin{aligned} 
\frac{\tau^2}{\tau'^2} &= \frac{1+k'}{1-k'}, & \frac{\tau^2-\tau'^2}{\tau^2+\tau'^2} &= k'. 
\end{aligned} 
\end{align}

According to (\ref{eqn:113}) and (\ref{eqn:32}) it now is
\begin{align}	\label{eqn:118}
2 \varrho = e^{-\alpha \frac{t}{\tau}} \cdot \cos \frac{t}{\tau} - e^{-\alpha' \frac{t}{\tau'}} \cdot \cos \frac{t}{\tau'}, 
\end{align}
whereas it is set
\begin{align}	\label{eqn:119}
\begin{aligned} 
\alpha &= \frac{\gamma_1+\gamma_2}{2} \cdot \frac{1}{2 \pi \sqrt{1+k'}}, & \alpha' &= \frac{\gamma_1+\gamma_2}{2} \cdot \frac{1}{2 \pi \sqrt{1-k'}}. 
\end{aligned} 
\end{align}
For numerical calculation it is more comfortable to write (\ref{eqn:118}) in the form
\begin{align}	\label{eqn:120}
2 \varrho = e^{-\alpha \frac{t}{\tau}} \cdot \cos \frac{t}{\tau} - e^{-\alpha'' \frac{t}{\tau'}} \cdot \cos \frac{t}{\tau'}, 
\end{align}
whereas it is
\begin{align}	\label{eqn:121}
\alpha'' = \alpha \cdot \frac{\tau^2}{\tau'^2}. 
\end{align}

The dependency of $\varrho$ on the ratio $\tau:\tau'$ and the middle logarithmic decrement $\frac{1}{2}(\gamma_1+\gamma_2)$  can only be obtained clearly with numerical evaluation. The following tables give the successive maxima of $\varrho$ and the values $t/T_1$ at which they occur. This is given for various ratios $T:T'=\tau:\tau'$ and for various logarithmic decrements $\gamma_1+\gamma_2/2$ how they can appear in practice.\footnote{The decrement $\gamma=0.15$ is actually pretty small. It has been observed by me while avoiding Hertzacher radiation in Abhandl.~d.~s\"{a}chs.~Gesellsch.~d~Wissensch.~{\bf 23.}~p.~99.~1896. --- One still can produce smaller decrements though (e.g.~$\gamma_1=0.09$) which I will report about later.} The absolute maxima are printed in bold. $T_1$ is the (common) oscillation period of the primary or secondary circuit, which every circuit possesses without coupling. The maxima of $2\varrho$ which follow the last maximum given in the tables will be always smaller than the last one stated.

\begin{center}	
\rule[0.3em]{0.3 \columnwidth}{0.5pt}
{\bf Page 546} 
\rule[0.3em]{0.3 \columnwidth}{0.5pt}
\end{center}

\begin{center}
\begin{tabular}{ c | c | c || c | c | c }
\multicolumn{6}{c}{$T:T' = 1.1, \quad k' = 0.095, \quad k'^2 = 0.0090. $} \\
\hline \hline 
\multirow{2}{*}{No.~of Max.} & \multicolumn{2}{ c || }{$\frac{\gamma_1+\gamma_2}{2} = 0.15$} & \multirow{2}{*}{No.~of Max.} & \multicolumn{2}{ c }{$\vphantom{\bigg|}\frac{\gamma_1+\gamma_2}{2} = 0.15$} \\
\cline{2-3} \cline{5-6} 
 & $t:T_1$ & $2\varrho$ &  & $t:T_1$ & $2\varrho$ \\
\hline \hline 
1 & 0.34 & $+$0.17 
	& 11 & 5.23 & $+$1.12 \\
2 & 0.78 & $-$0.42 & \ldots & \ldots & \ldots \\ 
3 & 1.25 & $+$0.63 & \ldots & \ldots & \ldots \\ 
4 & 1.75 & $-$0.83 
	& 21* & 10.45 & $+$0.10 \\ 
5 & 2.24 & $+$0.97 & \ldots & \ldots & \ldots \\ 
6 & 2.74 & $-$1.07 
	& 27 & 13.66 & $+$0.36 \\ 
7 & 3.23 & $+$1.16 
	& 28 & 14.18 & $-$0.37 \\
8 & 3.73 & {\bf $-$1.18} 
	& 29 & 14.69 & $+$0.36 \\
9 & 4.23 & $+$1.18 
	& 30 & 15.19 & $-$0.35 \\
10 & 4.73 & $-$1.15 
	& 31 & 15.68 & $+$0.34 \\ 
& & & \ldots & \ldots & \ldots 
\end{tabular} \\
* Compare to the comment of the next table at maximum No.~9 * 
\begin{tabular}{ c || c | c || c | c || c | c || c | c || c | c }
\multicolumn{11}{c}{$T:T' = 1.25, \quad k' = 0.219, \quad k'^2 = 0.048.$} \\
\hline \hline 
\multirow{2}{*}{No.~of Max.} 
	& \multicolumn{2}{ c || }{$\frac{\gamma_1+\gamma_2}{2} = 0.15$} 
	& \multicolumn{2}{ c || }{$\frac{\gamma_1+\gamma_2}{2} = 0.25$} 
	& \multicolumn{2}{ c || }{$\frac{\gamma_1+\gamma_2}{2} = 0.5$} 
	& \multicolumn{2}{ c || }{$\frac{\gamma_1+\gamma_2}{2} = 0.75$} 
	& \multicolumn{2}{ c }{$\vphantom{\bigg|}\frac{\gamma_1+\gamma_2}{2} = 1.0$} \\ 
\cline{2-11} 
 & $t:T_1$ & $2\varrho$ & $t:T_1$ & $2\varrho$ & $t:T_1$ & $2\varrho$ & $t:T_1$ & $2\varrho$ & $t:T_1$ & $2\varrho$ \\
\hline \hline 
1 & 0.31 & $+$0.37 & 0.30 & $+$0.36 & 0.29 & $+$0.32 & 0.27 & $+$0.29 & 0.26 & $+$0.27 \\
2 & 0.75 & $-$0.90 & 0.74 & $-$0.83 & 0.73 & $-$0.68 & 0.71 & $-$0.57 & 0.70 & {\bf $-$0.48} \\
3 & 1.23 & $+$1.26 & 1.22 & $+$1.11 & 1.21 & {\bf $+$0.82} & 1.19 & {\bf $+$0.61} & 1.18 & $+$0.46 \\
4 & 1.71 & {\bf $-$1.44} & 1.71 & {\bf $-$1.20} & 1.70 & $-$0.78 & 1.69 & $-$0.51 & 1.67 & $-$0.35 \\
5 & 2.20 & $+$1.42 & 2.20 & $+$1.13 & 2.20 & $+$0.65 & 2.19 & $+$0.38 & 2.18 & $+$0.23 \\
6 & 2.69 & $-$1.23 & 2.69 & $-$0.94 & 2.70 & $-$0.48 & 2.70 & $-$0.25 & 2.70 & $-$0.13 \\
7 & 3.18 & $+$0.93 & 3.18 & $+$0.67 & 3.19 & $+$0.31 & 3.20 & $+$0.15 & 3.22 & $+$0.07 \\
8 & 3.68 & $-$0.56 & 3.79 & $-$0.39 & 3.75 & $-$0.17 & 3.77 & $-$0.08 & 3.79 & $-$0.04 \\
9* & 4.41 & $+$0.15 & 4.41 & $+$0.16 & 4.41 & $+$0.11 & 4.41 & $+$0.05 & 4.41 & $+$0.02 \\
10 & 5.13 & $-$0.45 & 5.08 & $-$0.29 & 5.07 & $-$0.10 & 4.99 & $-$0.05 & --- & --- \\
11 & 5.62 & $+$0.64 & 5.61 & $+$0.38 & 5.58 & $+$0.11 & 5.54 & $+$0.03 & --- & --- \\
12 & 6.11 & $-$0.73 & 6.10 & $-$0.40 & 6.08 & $-$0.10 & 6.07 & $-$0.02 & --- & --- \\
13 & 6.62 & $+$0.73 & 6.62 & $+$0.38 & 6.62 & $+$0.09 & 6.62 & $+$0.02 & --- & --- 
\end{tabular} 
\end{center}

* No.~9 is no maximum for small $\frac{\gamma_1+\gamma_2}{2}$ (for $\frac{\gamma_1+\gamma_2}{2} = 0.15$ and $0.25$) but a minimum. Between the eighth and the tenth maximum, $\varrho$ namely takes then two positive maxima and a positive minimum. The latter is rubricated under No.~$9^*.$ For larger $\frac{\gamma_1+\gamma_2}{2}$ $9^*$ is a real maximum though, since in between 8 and 10 there is then only this one maximum.

\begin{center}	
\rule[0.3em]{0.3 \columnwidth}{0.5pt}
{\bf Page 547} 
\rule[0.3em]{0.3 \columnwidth}{0.5pt}
\end{center}

\begin{center}
\begin{tabular}{ c || c | c || c | c || c | c || c | c || c | c }
\multicolumn{11}{c}{$T:T' = 1.29, \quad k' = 0.249, \quad k'^2 = 0.0062. $} \\
\hline \hline 
\multirow{2}{*}{No.~of Max.} 
	& \multicolumn{2}{ c || }{$\frac{\gamma_1+\gamma_2}{2} = 0.15$} 
	& \multicolumn{2}{ c || }{$\frac{\gamma_1+\gamma_2}{2} = 0.25$} 
	& \multicolumn{2}{ c || }{$\frac{\gamma_1+\gamma_2}{2} = 0.5$} 
	& \multicolumn{2}{ c || }{$\frac{\gamma_1+\gamma_2}{2} = 0.75$} 
	& \multicolumn{2}{ c }{$\vphantom{\bigg|}\frac{\gamma_1+\gamma_2}{2} = 1.0$} \\ 
\cline{2-11} 
 & $t:T_1$ & $2\varrho$ & $t:T_1$ & $2\varrho$ & $t:T_1$ & $2\varrho$ & $t:T_1$ & $2\varrho$ & $t:T_1$ & $2\varrho$ \\
\hline \hline 
1 & 0.31  & $+$0.42 & 0.30 & $+$0.42 & 0.28 & $+$0.36 & 0.27 & $+$0.33 & 0.26 & $+$0.30 \\
2 & 0.75 & $-$1.01 & 0.74 & $-$0.93 & 0.73 & $-$0.77 & 0.71 & $-$0.64 & 0.70 & {\bf $-$0.54} \\ 
3 & 1.22 & $+$1.39 & 1.21 & $+$1.23 & 1.21 & {\bf $+$0.89} & 1.19 & {\bf $+$0.66} & 1.18 & $+$0.50 \\
4 & 1.71 & {\bf $-$1.50} & 1.70 & {\bf $-$1.26} & 1.70 & $-$0.82 & 1.69 & $-$0.54 & 1.68 & $-$0.36 \\
5 & 2.20 & $+$1.38 & 2.20 & $+$1.09 & 2.19 & $+$0.63 & 2.19 & $+$0.37 & 2.18 & $+$0.22 \\ 
\multicolumn{11}{c}{ } \\
\multicolumn{11}{c}{$T:T' = 1.332, \quad k' = 0.279, \quad k'^2 = 0.078.$} \\
\hline \hline
1 & 0.31 & $+$0.48 & 0.30 & $+$0.46 & 0.28 & $+$0.41 & 0.27 & $+$0.37 & 0.25 & $+$0.34 \\
2 & 0.75 & $-$1.11 & 0.74 & $-$1.03 & 0.72 & $-$0.85 & 0.71 & $-$0.71 & 0.69 & {\bf $-$0.59} \\
3 & 1.22 & $+$1.49 & 1.21 & {\bf $+$1.31} & 1.20 & {\bf $+$0.95} & 1.19 & {\bf $+$0.71} & 1.16 & $+$0.52 \\
4 & 1.69 & {\bf $-$1.52} & 1.69 & $-$1.28 & 1.69 & $-$0.82 & 1.69 & $-$0.54 & 1.68 & $-$0.36 \\ 
\multicolumn{11}{c}{ } \\
\multicolumn{11}{c}{$T:T' = 1.412, \quad k' = 0.332, \quad k'^2 = 0.110.$} \\
\hline \hline
1 & 0.30 & $+$0.56 & 0.29 & $+$0.54 & 0.28 & $+$0.49 & 0.27 & $+$0.45 & 0.25 & $+$0.41 \\ 
2 & 0.73 & $-$1.28 & 0.72 & $-$1.18 & 0.72 & $-$0.97 & 0.70 & {\bf $-$0.81} & 0.68 & {\bf $-$0.69} \\ 
3 & 1.19 & {\bf $+$1.59} & 1.19 & {\bf $+$1.40} & 1.18 & {\bf $+$1.03} & 1.17 & $+$0.77 & 1.15 & $+$0.57 \\ 
\multicolumn{11}{c}{ } \\
\multicolumn{11}{c}{$T:T' = 1.50, \quad k' = 0.384, \quad k'^2 = 0.148.$} \\
\hline \hline
1 & 0.30 & $+$0.66 & 0.29 & $+$0.63 & 0.27 & $+$0.57 & 0.26 & $+$0.51 & 0.25 & $+$0.48 \\
2 & 0.72 & $-$1.44 & 0.71 & $-$1.32 & 0.69 & {\bf $-$1.09} & 0.68 & {\bf $-$0.91} & 0.67 & {\bf $-$0.77} \\ 
3 & 1.17 & {\bf $+$1.63{\scriptsize 0}} & 1.17 & {\bf $+$1.44} & 1.17 & $+$1.04 & 1.16 & $+$0.77 & 1.15 & $+$0.58 \\
4 & 1.64 & $-$1.22 & 1.65 & $-$1.01 & 1.66 & $-$0.66 & 1.67 & $-$0.45 & 1.68 & $-$0.31 \\ 
5* & 2.35 & $+$0.22 & 2.35 & $+$0.26 & 2.35 & $+$0.28 & 2.35 & $+$0.22 & 2.35 & $+$0.16 \\
6 & 3.05 & $-$0.97 & 3.04 & $-$0.71 & 3.01 & $-$0.36 & 2.97 & $-$0.21 & 2.96 & $-$.012 \\ 
7 & 3.53 & $+$1.10 & 3.52 & $+$0.77 & 3.52 & $+$0.34 & 3.52 & $+$0.16 & 3.51 & $+$0.08 \\ 
\multicolumn{11}{c}{ } \\
\multicolumn{11}{c}{$T:T' = 1.667, \quad k' = 0.470, \quad k'^2 = 0.221.$} \\
\hline \hline
1 & 0.28 & $+$0.80 & 0.27 & $+$0.76 & 0.26 & $+$0.69 & 0.25 & $+$0.62 & 0.20 & $+$0.58 \\ 
2 & 0.69 & {\bf $-$1.61{\scriptsize 5}} & 0.68 & {\bf $-$1.49} & 0.67 & {\bf $-$1.21} & 0.66 & {\bf $-$1.01} & 0.65 & {\bf $-$0.85} \\ 
3 & 1.13 & $+$1.49 & 1.13 & $+$1.31 & 1.13 & $+$0.94 & 1.13 & $+$0.70 & 1.14 & $+$0.53 \\ 
4* & 1.82 & $-$0.24 & 1.82 & $-$0.31 & 1.82 & $-$0.36 & 1.82 & $-$0.32 & 1.82 & $-$0.26 \\
5 & 2.50 & $+$1.17 & 2.49 & $+$0.90 & 2.46 & $+$0.50 & 2.44 & $+$0.31 & 2.43 & $+$0.20 
\end{tabular} 
\end{center}

\begin{center}	
\rule[0.3em]{0.3 \columnwidth}{0.5pt}
{\bf Page 548} 
\rule[0.3em]{0.3 \columnwidth}{0.5pt}
\end{center}

\begin{center}
\begin{tabular}{ c || c | c || c | c || c | c || c | c || c | c }
\multicolumn{11}{c}{$T:T' = 2.0, \quad k' = 0.60, \quad k'^2 = 0.360.$} \\
\hline \hline 
\multirow{2}{*}{No.~of Max.} 
	& \multicolumn{2}{ c || }{$\frac{\gamma_1+\gamma_2}{2} = 0.15$} 
	& \multicolumn{2}{ c || }{$\frac{\gamma_1+\gamma_2}{2} = 0.25$} 
	& \multicolumn{2}{ c || }{$\frac{\gamma_1+\gamma_2}{2} = 0.5$} 
	& \multicolumn{2}{ c || }{$\frac{\gamma_1+\gamma_2}{2} = 0.75$} 
	& \multicolumn{2}{ c }{$\vphantom{\bigg|}\frac{\gamma_1+\gamma_2}{2} = 1.0$} \\ 
\cline{2-11} 
 & $t:T_1$ & $2\varrho$ & $t:T_1$ & $2\varrho$ & $t:T_1$ & $2\varrho$ & $t:T_1$ & $2\varrho$ & $t:T_1$ & $2\varrho$ \\
\hline \hline 
1 & 0.25 & $+$1.02 & 0.25 & $+$0.97 & 0.24 & $+$0.88 & 0.23 & $+$0.80 & 0.21 & $+$0.74 \\ 
2 & 0.63 & {\bf $-$1.74} & 0.63 & {\bf $-$1.60} & 0.62 & {\bf $-$1.28} & 0.62 & {\bf $-$1.04} & 0.61 & {\bf $-$0.89} \\ 
3* & 1.26 & $+$0.28 & 1.26 & $+$0.36 & 1.26 & $+$0.48 & 1.26 & $+$0.46 & 1.26 & $+$0.42 \\
4 & 1.90 & $-$1.33 & 1.89 & $-$1.05 & 1.88 & $-$0.64 & 1.87 & $-$0.44 & 1.86 & $-$0.31 \\ 
\multicolumn{11}{c}{ } \\
\multicolumn{11}{c}{$T:T' = 2.5, \quad k'= 0.724, \quad k'^2 = 0.524.$} \\
\hline \hline
1 & 0.22 & $+$1.27 & 0.22 & $+$1.20 & 0.21 & $+$1.06 & 0.18 & {\bf $+$0.96} & 0.18 & {\bf $+$0.88} \\ 
2 & 0.54 & {\bf $-$1.55} & 0.54 & {\bf $-$1.41} & 0.55 & {\bf $-$1.10} & 0.55 & $-$0.91 & 0.56 & $-$0.76 \\ 
3 & 1.31 & $+$1.39 & 1.31 & $+$1.14 & 1.31 & $+$0.77 & 1.31 & $+$0.59 & 1.31 & $+$0.49 \\ 
\multicolumn{11}{c}{ } \\
\multicolumn{11}{c}{$T:T' = 3, \quad k'= 0.80, \quad k'^2 = 0.64.$} \\
\hline \hline
1 & 0.19 & {\bf $+$1.39} & 0.19 & {\bf $+$1.32} & 0.18 & {\bf $+$1.18} & 0.18 & {\bf $+$1.04} & 0.17 & {\bf $+$0.95} \\ 
2* & 0.67 & $-$0.36 & 0.67 & $-$0.48 & 0.67 & $-$0.65 & 0.67 & $-$0.68 & 0.67 & $+$0.65 \\ 
3* & 1.34 & $+$0.54 & 1.34 & $+$0.65 & 1.34 & $+$0.65 & 1.34 & $+$0.57 & 1.34 & $+$0.49 \\ 
4* & 2.01 & $-$0.64 & 2.01 & $-$0.68 & 2.01 & $-$0.57 & 2.01 & $-$0.45 & 2.01 & $-$0.34 \\ 
5* & 2.68 & $+$0.66 & 2.68 & $+$0.66 & 2.68 & $+$0.49 & 2.68 & $+$0.35 & 2.68 & $+$0.24 \\ 
\multicolumn{11}{c}{ } \\
\multicolumn{11}{c}{$T:T' = 4, \quad k'= 0.882, \quad k'^2 = 0.778.$} \\
\hline \hline
1 & 0.15 & {\bf $+$1.52} & 0.15 & {\bf $+$1.42} & 0.14 & {\bf $+$1.21} & 0.13 & {\bf $+$1.08} & 0.12 & {\bf $+$1.01} \\ 
2 & 0.69 & $-$1.39 & 0.69 & $-$1.16 & 0.69 & $-$0.90 & 0.69 & $-$0.77 & 0.69 & $-$0.69 \\ 
3* & 1.37 & $+$0.73 & 1.37 & $+$0.82 & 1.37 & $+$0.69 & 1.37 & $+$0.58 & 1.37 & $+$0.49 \\ 
4 & 2.06 & $-$0.93 & 2.06 & $-$0.77 & 2.06 & $-$0.58 & 2.06 & $-$0.45 & 2.06 & $-$0.34 \\ 
5* & 2.74 & $+$0.77 & 2.74 & $+$0.69 & 2.74 & $+$0.49 & 2.74 & $+$0.35 & 2.74 & $-$0.24 \\ 
\multicolumn{11}{c}{ } \\
\multicolumn{11}{c}{$T:T' = 5, \quad k'= 0.923, \quad k'^2 = 0.854.$} \\
\hline \hline
1 & 0.12 & {\bf $+$1.56} & 0.11 & {\bf $+$1.46} & 0.11 & {\bf $+$1.27} & 0.10 & {\bf $+$1.12} & 0.10 & {\bf $+$1.02} \\ 
2* & 0.69 & $-$0.69 & 0.69 & $-$0.80 & 0.69 & $-$0.83 & 0.69 & $-$0.76 & 0.69 & $-$0.70 \\
3* & 1.39 & $+$0.83 & 1.39 & $+$0.82 & 1.39 & $+$0.70 & 1.39 & $+$0.58 & 1.39 & $+$0.49 \\
4* & 2.08 & $-$0.83 & 2.08 & $-$0.76 & 2.08 & $-$0.58 & 2.08 & $-$0.45 & 2.08 & $-$0.34 \\
5* & 2.77 & $+$0.81 & 2.77 & $+$0.69 & 2.77 & $+$0.49 & 2.77 & $+$0.35 & 2.77 & $+$0.24 \\ 
6* & 3.47 & $-$0.77 & 3.47 & $-$0.63 & 3.47 & $-$0.41 & 3.47 & $-$0.26 & 3.47 & $-$0.17 \\ 
\multicolumn{11}{c}{ } \\
\multicolumn{11}{c}{$T:T' = 10, \quad k'= 0.98, \quad k'^2 = 0.96.$} 
\end{tabular} 
\end{center}

\noindent
1. Max.~(likewise absolute) at $\frac{\gamma_1+\gamma_2}{2}=0.15$ for $t/T_1 = 0.066$: ${\bf 2\boldsymbol{\varrho} = +1.55.}$ 
\begin{align} \nonumber 
\begin{aligned}
T:T' &= 100, & k' &= 0.9998. 
\end{aligned}
\end{align}
1. Max.~(likewise absolute) at $\frac{\gamma_1+\gamma_2}{2}=0.15$ for $t/T_1 = 0.0047$: ${\bf 2 \boldsymbol{\varrho} = +1.02.}$ 

\begin{center}	
\rule[0.3em]{0.3 \columnwidth}{0.5pt}
{\bf Page 549} 
\rule[0.3em]{0.3 \columnwidth}{0.5pt}
\end{center}

The maxima indicated with $*$ are no maxima for small $\gamma_1+\gamma_2/2$, but minima of absolute value $2\varrho$ (see comment about p.~546). E.g.~it is for 
\begin{align}	\nonumber
\begin{aligned}
T:T' &= 3, & \frac{\gamma_1+\gamma_2}{2} &= 0.15, & t:t_1 &= 0.45, & 2 \varrho &= -1.22. \\
T:T' &= 2, & \frac{\gamma_1+\gamma_2}{2} &= 0.15, & t:t_1 &= 1.02, & 2 \varrho &= +0.82. \\
T:T' &= 1.5, & \frac{\gamma_1+\gamma_2}{2} &= 0.15, & t:t_1 &= 2.09, & 2 \varrho &= +0.44. 
\end{aligned}
\end{align}
The behavior of $2 \varrho$ is obtained from the graphical illustrations (see below). 

First, we are interested in the absolute maxima printed in bold in the tables. One notices, that generally the same ones increase with increasing coupling and decreasing damping, \textit{but for not too large damping ($\gamma_1+\gamma_2/2 \leq 0.5$), the coupling\footnote{I always only give the coefficient $k'$ or $k'^2$ here. The actual coupling coefficient $k^2$ can be calculated from $k'^2$ using equation (\ref{eqn:115}), if $\gamma_1-\gamma_2$ is known. For $k'^2=0.36$ the difference between $k'^2$ and $k^2$ can already be neglected.} $k'=0.6$, $k'^2=0.36$, $T:T'=2$ is most convenient, i.e.~it is also better than stronger couplings.} The dependencies of the absolute maxima $2\bar{\varrho}$ from the coupling is shown in Fig.~2, in which the coefficient $k'$ and the absolute maximum $2\bar{\varrho}$ are plotted as ordinates. 

The curves have a complicated character. For the coupling of $k'=0.6$ and $\gamma_1+\gamma_2/2=0.15$, it is $2\bar{\varrho}=1.74$, i.e.~it is 13 percent less than the value $2\bar{\varrho}=2$, which would occur for vanishing damping. 

For $\gamma_1+\gamma_2/2=1$ and $k'=0.6$ is it $2\bar{\varrho}=0.89$, i.e.~it is about half as large as $2\bar{\varrho}$ for $\gamma_1+\gamma_2/2=0.15$. \textit{Hence, with such strong coupling the maximum amplitude depends only little on the damping, however, it does substantially with small coupling.} From Fig.~2 one gets the best idea of the borders, in which the efficiency can vary due to changing damping, for a Tesla coil built with good resonance. Here, it has also to be taken into account that following (\ref{eqn:112}) due to the factor
\begin{align}	\nonumber
\sqrt{k^2:k^2-\left(\frac{\gamma_1-\gamma_2}{2 \pi}\right)^2} = \sqrt{k'^2+\left(\frac{\gamma_1-\gamma_2}{2 \pi}\right)^2:k'^2}
\end{align}

\begin{center}	
\rule[0.3em]{0.3 \columnwidth}{0.5pt}
{\bf Page 550} 
\rule[0.3em]{0.3 \columnwidth}{0.5pt}
\end{center}

\noindent 
the maximum of $V^2$ with small $k'^2$ gets even larger than 
\begin{align} \nonumber
\frac{\varrho}{4} F \sqrt{\frac{C_1}{C_2} \frac{L_{21}}{L_{12}} }, 
\end{align}
namely when $(\gamma_1-\gamma_2/2\pi)^2$ has a noticeable value against $k'^2.$ 
 
\begin{figure}[h!]	
\begin{center}	
	\includegraphics[width=0.6\columnwidth]{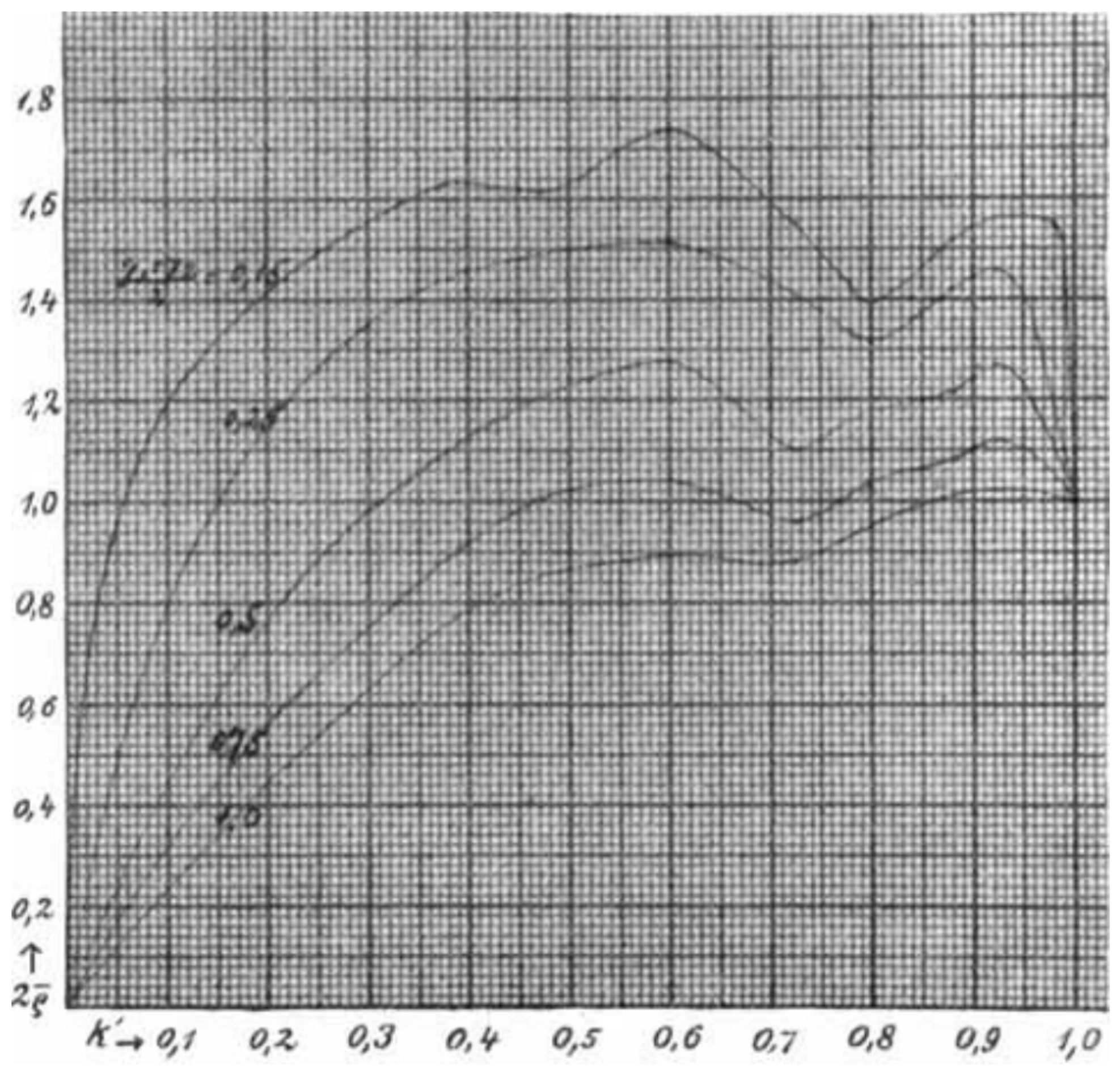}
	\caption{\label{fig1}Reproduction of Fig.~2 on p.~550 of Drude~\cite{drude:1904}. Copyright Wiley-VCH Verlag GmbH \& Co.~KGaA. Reproduced with permission.  } \addcontentsline{toc}{subsubsection}{Figure 2}
\end{center}
\end{figure}

\begin{center}
\textbf{VII. Application to Wireless Telegraphy.}\addcontentsline{toc}{subsubsection}{VII.~Application to Wireless Telegraphy}
\end{center}
 
When using a sending apparatus with inductive excitement (magnetic coupling), formula (\ref{eqn:112}) essentially still stands, despite that the secondary circuit (Tesla coil) is provided with antennas to increase the radiation. When now waiting on resonance in the receiving apparatus, it then only depends on the gain of a high as possible absolute maximum of $V^2.$ Hence, one will choose the coupling $k'=0.6$ (only with very high damping $\gamma_1+\gamma_2/2  \geq 0.75$ one would use a even higher one if possible). When using the resonance principle though 

\begin{center}	
\rule[0.3em]{0.3 \columnwidth}{0.5pt}
{\bf Page 551} 
\rule[0.3em]{0.3 \columnwidth}{0.5pt}
\end{center}

\noindent
in order to adjust the receiver to the sender, the temporal progression of the factor $\varrho$ in formula (\ref{eqn:112}) is important. One has to differentiate between two cases now, depending on if one is using a onefold uncoupled or only very loosely coupled (twofold) receiving apparatus, which has only \textit{one} eigenoscillation period, or if one is using a tightly coupled twofold receiver, which holds \textit{two} different eigenperiods. 

\begin{figure}[h]	
\begin{center}	
	\includegraphics[width=0.6\columnwidth]{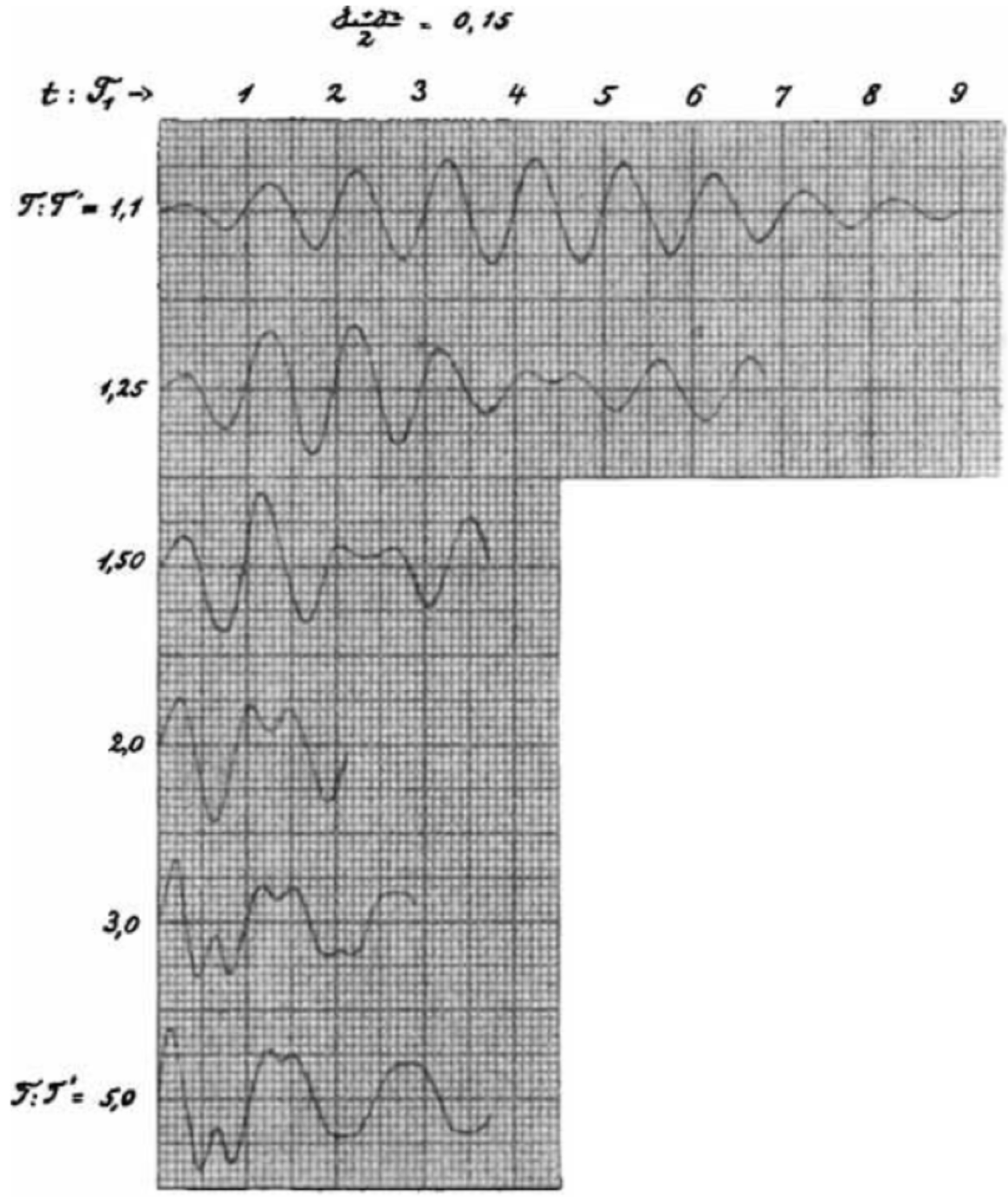}
	\caption{\label{fig3}Reproduction of Fig.~3 on p.~551 of Drude~\cite{drude:1904}. Copyright Wiley-VCH Verlag GmbH \& Co.~KGaA. Reproduced with permission.  } \addcontentsline{toc}{subsubsection}{Figure 3}
\end{center}
\end{figure}

a)  \textit{Onefold or loose coupled twofold receiving apparatus.} If the apparatus holds only \textit{one} eigenperiod, in order to receive maximal efficiency, $\varrho$ has to have a progression as periodical as possible and the receiver's eigenperiod has to concur with this period of $\varrho$. 
 
The temporal progression of $\varrho$ is gained roughly from the maxima which are shown in the tables, and a little better (due to the special shaping of the $\varrho-$curve in the surrounding of the maxima indicated with *) from the graphical representation of Figure 3. This figure displays the temporal progression of $\varrho$ with various $k'$ only for the smallest damping $\gamma_1+\gamma_2/2=0.15$, because here the characteristics manifest the most. 
 
The general character of these curves, especially with small coupling ($T:T'=1.1$ and $T:T'=1.25$) is the one of the so called \textit{beats}. The first beat is terminated at the maxima indicated with * in the tables (in the curves recognizable as a saddle). Within a beat, 

\begin{center}	
\rule[0.3em]{0.3 \columnwidth}{0.5pt}
{\bf Page 552} 
\rule[0.3em]{0.3 \columnwidth}{0.5pt}
\end{center}

\noindent 
the progression of $\varrho$ is nearly periodic with the period $\frac{1}{2} T_1$, but the second beat is phase shifted to the first by $\pi$, so that on a receiving apparatus, which is built as a onefold resonator with period $T_1$, \textit{the second beat is harmful for the effect of the first beat.} Therefore, if, for wireless telegraphy, one place emphasis on tuning, one has to choose the sender's coupling so that the second beat receives a maximal amplitude which is relatively small in comparison to the maximal amplitude of the first beat. This is the more the case, the smaller the coupling is in the sending apparatus. This is because following the tables, e.g., for $\gamma_1+\gamma_2/2=0.15$ the maximal amplitudes $\bar{\varrho_1}$ and $\bar{\varrho_2}$ of the first two beats are:
\begin{align}
\begin{aligned} 
\text{At}& & T:T' &= 1.1, & k' &= 0.095, & 2 \bar{\varrho}_1 &= 1.18, & 2 \bar{\varrho}_2 &= 0.37, & \bar{\varrho}_1:\bar{\varrho}_2 &= 3.19. \nonumber \\ 
''& &  &= 1.25, &  &= 0.219, & &= 1.44, &  &= 0.73, &  &= 1.97. \nonumber \\ 
''& &  &= 1.50, &  &= 0.384, & &= 1.63, &  &= 1.10, &  &= 1.48. \nonumber 
\end{aligned} 
 \end{align}
 
Therefore, the coupling has to be very small in the sender when a sharp resonance in a onefold receiver is supposed to be achieved.\footnote{The result agrees with the one already derived from W.~Wien (Ann.~d.~Phys.~{\bf 8.}~p.~711.~1902). For small couplings there is actually no difference between Wien's and my derivations but only for very strong coupling. Following my derivations for very strong coupling good resonance would be possible, although it is to be doubted if it can be realized. (Compare further down in the text).} Now, if one also places emphasis on on highest possible sensitivity of the receiver though, it has to be taken into account, that the maximal amplitude of $\bar{\varrho_1}$ of the first beat decreases with decreasing coupling. For that reason one would not choose a small coupling. The best choice of the coupling in the sender is depending on the damping $\gamma_1+\gamma_2/2$ in the sender as well as the eigendamping of the receiver. It is obvious, that for a strongly damped receiver, whose eigenoscillation has almost vanished after 20 half eigenperiods ($\gamma=0.4$, amplitude only 2 percent of the start amplitude), the second beat of the coupling $T:T'=1.1, k'=0.095$, which just starts at after 21 half eigenperiods $T_1$ is totally unharmful. Thus,

\begin{center}	
\rule[0.3em]{0.3 \columnwidth}{0.5pt}
{\bf Page 553} 
\rule[0.3em]{0.3 \columnwidth}{0.5pt}
\end{center}

\noindent
for such a receiver one would definitely not choose an even smaller coupling in the sender. But also, the right choice of the coupling depends on the damping $\gamma_1+\gamma_2/2$ in the sender, since the ratio $\bar{\varrho_1}:\bar{\varrho_2}$ of the maximal amplitudes of the first both beats becomes higher, the higher $\gamma_1+\gamma_2/2$ is. This is shown by the following table: 
\begin{center}
\begin{tabular*}{\textwidth}{ c @{\extracolsep{\fill}} | c c c c c }
	\multicolumn{6}{c}{$T:T' = 1.25, \quad k' = 0.219.$} \\
\hline \hline
$\gamma_1+\gamma_2/2$~	& 0.15	& 0.25	& 0.5		& 0.75	& 1.0 \\
$2\bar{\varrho}_1$				& 1.44	& 1.20	& 0.82	& 0.61	& 0.48 \\
$2\bar{\varrho}_2$				& 0.73	& 0.40	& 0.11	& 0.05	& 0.02 \\
$\bar{\varrho}_1:\bar{\varrho}_2$	& 1.97	& 3.0		& 7.5		& 12.2	& 24.0 \\ 
	\multicolumn{6}{c}{$T:T' = 1.50, \quad k' = 0.384.$} \\
$\gamma_1+\gamma_2/2$~	& 0.15	& 0.25	& 0.5		& 0.75	& 1.0 \\
$2\bar{\varrho}_1$				& 1.63	& 1.44	& 1.09	& 0.91	& 0.77 \\
$2\bar{\varrho}_2$				& 1.10	& 0.77	& 0.36	& 0.22	& 0.16 \\
$\bar{\varrho}_1:\bar{\varrho}_2$	& 1.48	& 1.87	& 3.0		& 4.1		& 4.8 
\end{tabular*}
\end{center}

Thus, when, for example,  the second beat can be a tenth of the amplitude of the first beat without its destructive impact becoming noticeable and when it is $\gamma_1+\gamma_2/2=0.75$ in the sender, one does not need to choose a weaker coupling than $k'=0.219$, since already for this coupling it is $\bar{\varrho}_1:\bar{\varrho}_2=12.2$. 

When we now investigate the bottom curves of Fig.~3, which are valid for a very strong coupling, again different ratios become valid, since $\varrho$ becomes periodic without beat. This is clearly visible from the curves of Fig.~3 for ratios $T:T'$ larger than 3, i.e.~for couplings $k'$ larger than $k'=0.8$. The reason for this is that for very strong couplings the faster oscillation of the two periods $T$, $T'$ of the sender is more strongly damped than the slower one, which then leave only that one noticeable. \textit{Therefore, if couplings which are larger than $k'=0.8$ can be experimentally realized, for gaining better resonance and strong effectivity one can also work with very strong coupling in the sender. Therefore, the eigenperiod of the receiver has then to be $\sqrt{2}=1.41$ times larger than the eigenperiod $T_1$ of the two (uncoupled) oscillation circuits of the sender.}

\begin{center}	
\rule[0.3em]{0.3 \columnwidth}{0.5pt}
{\bf Page 554} 
\rule[0.3em]{0.3 \columnwidth}{0.5pt}
\end{center}

Following (\ref{eqn:60}), p.~524, \textit{the maximum amplitude of a simple receiver strongly depends of the damping of the sender and the receiver, and the integral effect depends on that even more, following (\ref{eqn:84})}. 

b)  \textit{Tightly coupled receiver} If the receiver has \textit{two} eigenperiods $T$ and $T'$ (due to the tight coupling of both of its parts) and when defining $V_2$ as the potential at the free end of the receiver's antennas and $V_1$ as the potential difference between the configuration of the capacitor in circuit 1 of the receiver which is tightly coupled with the circuit 2 of the receiver which contains the antennas, the following equations are valid according to equations (II) of p.~519:

\begin{align}	\label{eqn:122}
(\tau_1^2+\vartheta_1^2) \frac{d^2 V_1}{dt^2} + 2 \vartheta_1 \frac{d V_1}{dt} + V_1 &= p_{12}(\tau_1^2+\vartheta_1^2) \frac{d^2 V_2}{dt}, \\ 
(\tau_2^2+\vartheta_2^2) \frac{d^2 V_2}{dt^2} + 2 \vartheta_2 \frac{d V_2}{dt} + V_2 &= p_{21}(\tau_2^2+\vartheta_2^2) \frac{d^2 V_2}{dt} + a\cdot V_2'.  \label{eqn:123}
\end{align}

Here, now all values except $V_2'$ refer to the circuits 1 and 2 of the receiver. Only $V'_2$ refers to the sender and resembles the until now in this work used value $V_2$, i.e.~the potential at the end of the sender's antenna, for which there are more precise values given in formulas (\ref{eqn:112}) and (\ref{eqn:113}). The term $a V_2'$ resembles the excitement of the receiver (more precisely the receiver's antennas) via the electromagnetic force emitted by the waves of the sender. Therefore, the factor $a$ depends on the distance of the sender to the receiver and on the individual construction, e.g.~the length of the antenna of the receiver. 

Following (\ref{eqn:112}) and (\ref{eqn:113}) one has to set
\begin{align}	\label{eqn:124}
a V_2' = S\left( e^{-\frac{\delta t }{\tau^2}} \cos \frac{t}{\tau} - e^{-\frac{\delta t }{\tau'^2}} \cos \frac{t}{\tau'} \right). 
\end{align}

We can set: 
\begin{align}	\label{eqn:125}
a V_2' = S_1 e^{t/z_1} + S_2 e^{t/z_2} + S_3 e^{t/z_3} + S_4 e^{t/z_4}, 
\end{align}
whereas it is set
\begin{align}	\label{eqn:126}
\left\{ \begin{aligned}
z_1 &= - \delta + i \tau, & z_2 &= -\delta - i \tau, & z_3 &= - \delta + i \tau', \\ 
& & z_4 &= - \delta - i \tau', 
\end{aligned} \right.
\end{align}
and
\begin{align}	\label{eqn:127}
\begin{aligned} 
S_1 &= S_2 = S, & S_3 &= S_4 = -S. 
\end{aligned} 
\end{align}

\begin{center}	
\rule[0.3em]{0.3 \columnwidth}{0.5pt}
{\bf Page 555} 
\rule[0.3em]{0.3 \columnwidth}{0.5pt}
\end{center}

For the integral of (\ref{eqn:122}), (\ref{eqn:123}) we write analogously to the earlier formula (\ref{eqn:25}) p.~520:
\begin{align}	\label{eqn:128}
\left\{ \begin{aligned}
V_1 &= A_1 e^{t/y_1} + A_2 e^{t/y_2} +  A_3 e^{t/y_3} +  A_4 e^{t/y_4} \\
	&+ D_1 e^{t/z_1} + D_2 e^{t/z_2} + D_3 e^{t/z_3} + D_4 e^{t/z_4}, \\
V_2 &= B_1 e^{t/y_1} + B_2 e^{t/y_2} +  B_3 e^{t/y_3} +  B_4 e^{t/y_4} \\
	&+ E_1 e^{t/z_1} + E_2 e^{t/z_2} + E_3 e^{t/z_3} + E_4 e^{t/z_4}. \\
\end{aligned} \right.
\end{align}

The parts proportional to A and B are the eigenoscillations of the receiver, the parts proportional to D and E are the oscillations forced by the sender. Therefore, the $y$ are the four square roots of the equation (compare (\ref{eqn:29}) p.~520):
\begin{align}	\label{eqn:129}
\left\{ \begin{aligned}
(y^2 + 2 \vartheta_1 y + \tau_1^2 + \vartheta_1^2)(y_2 &+ 2 \vartheta_2 y + \tau_2^2 + \vartheta_2^2) \\
	&- p_{12}p_{21}(\tau_1^2 + \vartheta_1^2) (\tau_2^2 + \vartheta_2^2) \\
	= (y-y_1)(y-y_2)(y-&y_3)(y-y_4) = 0. 
\end{aligned} \right.
\end{align}

As initial condition (for $t=0$) it is:
\begin{align}	\nonumber
\begin{aligned}
V_1 &= 0, & V_2 &= 0, & i_1 &= 0, & i_2 &= 0, 
\end{aligned}
\end{align}
therefore it is (compare (\ref{eqn:102}) p.~537):
\begin{align}	\label{eqn:130}
\left\{ \begin{aligned}
\sum A + \sum D &= 0, & \sum B + \sum E &= 0, \\ 
\sum \frac{A}{y} + \sum \frac{D}{z} &= 0, & \sum \frac{B}{y} + \sum \frac{E}{z} &= 0. 
\end{aligned} \right.
\end{align}

Now, following (\ref{eqn:122}), it is:
\begin{align}	\label{eqn:131}
A_1(\tau_1^2 + \vartheta_1^2 + 2\vartheta_1 y_1 + y_1^2) &= p_{12}(\tau_1^2 + \vartheta_1^2) B_1, \\ 
D_1(\tau_1^2 + \vartheta_1^2 + 2\vartheta_1 z_1 + z_1^2) &= p_{12}(\tau_1^2 + \vartheta_1^2) E_1, \label{eqn:132}
\end{align}
and further six analogous equations for $A_2, y_2, B_2$ and $D_2,z_2,E_2$ etc. When dividing (\ref{eqn:131}) by $y_1$, (\ref{eqn:132}) by $z_1$ and then sum up the two, it yields:
\begin{align}	\nonumber
(\tau_1^2 + \vartheta_1^2)\left(\frac{A_1}{y_1} + \frac{D_1}{z_1} \right) &+ 2 \vartheta_1 (A_1 + D_1) + A_1 y_1 + D_1 z_1 \\
	&= p_{12} (\tau_1^2+\vartheta_1^2)\left(\frac{B_1}{y_1} + \frac{E_1}{z_1} \right). \nonumber 
\end{align}

When here adding the three analogous equations which one obtains when substituting the index 1 in $A,D,y,z$ with $2, 3, 4$, because of the relations (\ref{eqn:130}) it gets:
\begin{align}	\label{eqn:133}
\sum Z y + \sum D z = 0. 
\end{align}

When adding (\ref{eqn:131}) and (\ref{eqn:132}), it yields:
\begin{align}	\nonumber
(\tau_1^2+\vartheta_1^2) (A_1 + D_1) &+ 2 \vartheta_1 (A_1 y_1 + D_1 z_1) + A_1 y_1^2 + D_1 z_1^2 \\
	&=p_{12}(\tau_1^2 + \vartheta_1^2)(B_1+E_1). \nonumber 
\end{align}

\begin{center}	
\rule[0.3em]{0.3 \columnwidth}{0.5pt}
{\bf Page 556} 
\rule[0.3em]{0.3 \columnwidth}{0.5pt}
\end{center}

When additionally adding the three analogous equations which one gets when again substituting the index 1 in $A, D, y, z$ with $2, 3, 4$, due to (\ref{eqn:130}) and (\ref{eqn:133}) it gets: 
\begin{align}	\label{eqn:134}
\sum A y^2 + \sum D z^2 = 0. 
\end{align}

In (\ref{eqn:130}), (\ref{eqn:133}) and (\ref{eqn:134}) there now are four linear equations for $A_1, A_2, A_3, A_4$ present, with which one can express the $A$ with the $D$, e.g., it becomes:
\begin{align}	\nonumber
\frac{A_1}{y_1} \Delta = - y_2y_3y_4 \Delta_{11} \sum \frac{D}{z} + \Delta_{12} \sum D = D_{13} \sum D z + \Delta_{14} \sum D z^2. 
\end{align}

Here, the $\Delta$ are the determinants:
\begin{align}	\nonumber
\Delta &= 
\begin{vmatrix}
1, & 1, & 1, & 1 \\
y_1, & y_2, & y_3, & y_4 \\
y_1^2, & y_2^2, & y_3^2, & y_4^2 \\
y_1^3, & y_2^3, & y_3^3, & y_4^3 
\end{vmatrix} 
= (y_1 - y_2)(y_1-y_3)(y_1-y_4) (y_2 - y_3)(y_2-y_4)(y_3-y_4), \\ 
\Delta_{11} = \Delta_{14} &= 
\begin{vmatrix}
1, & 1, & 1 \\
y_2, & y_3, & y_4 \\
y_2^2, & y_3^2, & y_4^2
\end{vmatrix}  
= - (y_2-y_3)(y_2-y_4)(y_3-y_4), \nonumber \\
\Delta_{12} &=  
\begin{vmatrix}
1, & 1, & 1 \\
y_2^2, & y_3^2, & y_4^2 \\
y_2^3, & y_3^3, & y_4^3
\end{vmatrix}  
= - (y_2-y_3)(y_2-y_4)(y_3-y_4)(y_2y_3+y_2y_4+y_3y_4), \nonumber \\
\Delta_{13} &=
\begin{vmatrix}
1, & 1, & 1 \\
y_2, & y_3, & y_4 \\
y_2^3, & y_3^3, & y_4^3
\end{vmatrix}  
= - (y_2-y_3)(y_2-y_4)(y_3-y_4)(y_2 + y_3 + y_4).  \nonumber 
\end{align}

Therefore, it becomes
\begin{align}	\label{eqn:135}
\left\{ \begin{aligned}
\frac{A_1}{y_1}(y_1 &- y_2)(y_1-y_3)(y_1-y_4) = y_2y_3y_4 \sum \frac{D}{z} \\
	&- (y_2y_3 + y_2y_4+y_3y_4) \sum D \\
	&+ (y_2 + y_3+y_4)\sum D z - \sum D z^2 \\
	= &- \sum \frac{D}{z}(z-y_2)(z-y_3)(z-y_4). 
\end{aligned} \right.
\end{align}
Here, in the $\Sigma$ one has to take the indices $1, 2, 3, 4$ for $D$ and $z$ and add these four parts. From equation (\ref{eqn:123}) it now follows
\begin{align}	\nonumber
E_1(\tau_2^2+\vartheta_2^2 + 2 \vartheta_1 z_1 + z_1^2) = p_{21} (\tau_2^2+\vartheta_2^2) D_1 + S_1 z_1^2. 
\end{align}

When multiplying this equation with (\ref{eqn:132}) and dividing by $E_1$, it becomes:

\begin{center}	
\rule[0.3em]{0.3 \columnwidth}{0.5pt}
{\bf Page 557} 
\rule[0.3em]{0.3 \columnwidth}{0.5pt}
\end{center}

\begin{align}	\nonumber
D_1 [(z_1^2 &+ 2\vartheta_1z_1 + \tau_1^2 + \vartheta_1^2)(z_1^2+2\vartheta_2z_1 + \tau_2^2 + \vartheta_2^2) \\
	&- p_{12}p_{21} (\tau_1^2 + \vartheta_1^2)(\tau_2^2 + \vartheta_2^2)] = p_{12} (\tau_1^2 + \vartheta_1^2)S_1 z_1^2, \nonumber 
\end{align}
or due to (\ref{eqn:129}):
\begin{align}	\label{eqn:136}
D_1(z_1-y_1)(z_1-y_2)(z_1-y_3)(z_1-y_4) = p_{12} (\tau_1^2+\vartheta_1^2)z_1^2S_1. 
\end{align}
When plugging this value into (\ref{eqn:136}), it follows:
\begin{align}	\label{eqn:137}
\frac{A_1}{y_1} (z_1-y_1)(z_1-y_2)(z_1-y_3)(z_1-y_4) = - p_{12} (\tau_1^2+\vartheta_1^2)z_1^2 \sum S_n \frac{z_n}{z_n-y_1}. 
\end{align}
In the $\Sigma$-sign, $S$ and $z$ successively have the four indices $n=1,2,3,4$ while $y_1$ does not change its index. The formulas for $A_2, A_3, A_4$ result from (\ref{eqn:137}) by cyclic permutation of the indices 1, 2, 3, 4 of the values $A, S, y, z$ of formula (\ref{eqn:137}). 

The formulas (\ref{eqn:128}), (\ref{eqn:136}) and (\ref{eqn:137}) yield the full solution of the exercise for $V_1$ with arbitrary values $y$ and $z$, i.e.~with arbitrary damping and eigenoscillations of both sender and receiver. It is obvious that the receiver is excited the most in case of resonance; one therefore will arrange it so that the two eigenperiods of the sender are identical with the receiver. If, for example, identical constructed apparatuses (also in regards to the coupling) are used for sender and receiver with just the difference that the circuit 1 of the sender contains a spark gap while the receiver contains a current- or potential indicator at that place in the circuit instead (Rutherford's magnetic current indicator or coherer), the eigenperiods of the sender and the receiver are the same. In regards to (\ref{eqn:126}) one thus has to set:
\begin{align}	\label{eqn:138}
\left\{ \begin{aligned}
y_1 &= - \sigma + i \tau, & y_2 &= - \sigma - i \tau, \\ 
y_3 &= - \sigma + i \tau', & y_4 &= - \sigma - i \tau'. 
\end{aligned} \right.
\end{align} 

The damping value $\sigma$ of the eigenoscillation of the transmitter has generally to be assumed different than the damping value $\delta$ of the transmitter. For identically constructed apparatuses $\delta$ is usually larger than $\sigma$, since the sender contains a spark gap. But since the wave indicator in the receiver also consumes electrical energy it can possibly happen that $\sigma=\delta$. 
  
For further calculation we want to assume the case $\sigma=\delta$, since doing so makes the formulas very simple the essentials sufficiently apparent.

\begin{center}	
\rule[0.3em]{0.3 \columnwidth}{0.5pt}
{\bf Page 558} 
\rule[0.3em]{0.3 \columnwidth}{0.5pt}
\end{center}

Assuming $\delta=\sigma$, $y$ and $z$ become equal, i.e., it becomes $y_1=z_1, y_2=z_2, y_3=z_3, y_4=z_4$. We therefore can write, following (\ref{eqn:136}):
\begin{align}	\label{eqn:139}
D_1(y_1-y_2)(y_1-y_3)(y_1-y_4) = p_{12} (\tau_1^2+\vartheta_1^2) S_1 \frac{z_1^2}{z_1-y_1},  
\end{align}
and following (\ref{eqn:137}):
\begin{align}	\label{eqn:140}
A_1(y_1-y_2)(y_1-y_3)(y_1-y_4) = -p_{12} (\tau_1^2+\vartheta_1^2) S_1 \frac{z_1 y_1}{z_1-y_1}. 
\end{align}

Looking at the $\Sigma$ on the right side of (\ref{eqn:137}), namely, only the part with $S_1$ matters, since it contains the denominator $z_1-y_1=0$. 

From (\ref{eqn:128}) it further follows:
\begin{align}	\label{eqn:141} 
\left\{ \begin{aligned}
V_1 &= (A_1 + D_1) e^{t/y_1} + (A_2 + D_2) e^{t/y_2} \\ 
	&+ (A_3 + D_3) e^{t/y_3} + (A_4 + D_4) e^{t/y_4}.  
\end{aligned} \right.
\end{align}

With addition of (\ref{eqn:139}) and (\ref{eqn:140}) it yields:
\begin{align}	\nonumber
(A_1+D_1)(y_1-y_2)(y_1-y_3)(y_1-y_4) = p_{12}(\tau_1^2+\vartheta_1^2) S_1 y_1, 
\end{align}
or following (\ref{eqn:138}):
\begin{align}	\nonumber
(A_1+D_1)(\tau^2 - \tau'^2) = -\frac{1}{2} p_{12}(\tau_1^2+\vartheta_1^2)S_1(1+i\sigma/\tau) 
\end{align}
and analogously:
\begin{align}	\nonumber
(A_2+D_2)(\tau^2 - \tau'^2) &= -\frac{1}{2} p_{12}(\tau_1^2+\vartheta_1^2)S_2(1-i\sigma/\tau), \\ 
(A_3+D_3)(\tau^2 - \tau'^2) &= -\frac{1}{2} p_{12}(\tau_1^2+\vartheta_1^2)S_3(1+i\sigma/\tau'), \nonumber \\ 
(A_4+D_4)(\tau^2 - \tau'^2) &= -\frac{1}{2} p_{12}(\tau_1^2+\vartheta_1^2)S_4(1-i\sigma/\tau'). \nonumber 
\end{align}

With respect to (\ref{eqn:127}) and following (\ref{eqn:141}) and when neglecting $\vartheta^2_1$ against $\tau_1^2$ and $\sigma^2$ against $\tau^2$ respectively against $\tau'^2$:
\begin{align}	\label{eqn:142}
V_1 = - \frac{p_{12} \tau_1^2 S}{\tau^2 - \tau'^2} \left( e^{-\frac{\sigma t}{\tau^2}} \cdot \cos \frac{t-\sigma}{\tau} - e^{-\frac{\sigma t}{\tau'^2}} \cdot \cos \frac{t \sigma}{\tau'} \right).  
\end{align}

When setting $t-\sigma=t'$ and disregarding start times which are on the order of magnitude of $\tau=\sigma$, one can write equation (\ref{eqn:142}) as:
\begin{align}	\nonumber
V_1 = - \frac{p_{12} \tau_1^2 S}{\tau^2 - \tau'^2} \left( e^{-\frac{\sigma t'}{\tau^2}} \cdot \cos \frac{t'}{\tau} - e^{-\frac{\sigma t'}{\tau'^2}} \cdot \cos \frac{t'}{\tau'} \right).  
\end{align}

Following the expansions in section V p.~540 this has now to be rearranged to:
\begin{align}	\label{eqn:143}
V_1 = - \frac{S}{2 \pi} \sqrt{\frac{C_1}{C_2} \frac{L_{12}}{L_{21}} \cdot \frac{k^2}{k^2 - \left(\frac{\gamma_1-\gamma_2}{2 \pi}\right)^2 }} \left( e^{-\frac{\sigma t'}{\tau^2}} \cdot \cos \frac{t'}{\tau} - e^{-\frac{\sigma t'}{\tau'^2}} \cdot \cos \frac{t}{\tau'} \right).
\end{align}

\begin{center}	
\rule[0.3em]{0.3 \columnwidth}{0.5pt}
{\bf Page 559} 
\rule[0.3em]{0.3 \columnwidth}{0.5pt}
\end{center}

Hence, the potential difference $V_1$ in the receiver is expressed with an analogous formula like the potential $V_2$ at the transmitter following (\ref{eqn:112}), with the only difference being the change of roles of $C_1L_{21}$ and $C_2L_{12}$ as numerator and denominator. If $V_1$ should become as large as possible, one therefore has to choose $C_1$ small against $V_2$, but if one wants to receive the strongest primary current as possible $C_1$ has to be large against $C_2$, since it is
\begin{align}	\nonumber
i_1 = - C_1 \frac{d V_1}{dt} 
\end{align}
following (\ref{eqn:1}) p.~514. At least, one has to choose this arrangement for Rutherford's magnetic wave indicator and it seems to me that even for the coherer, due to it's large and inconstant capacity, this arrangement is more practical than the recently mostly chosen transformation to a high potential $V_1$. At least when one wants to work with a distinct resonance.\footnote{I will  discuss this point in more detail at a later occasion.} 
 
The dependency of $V_1$ on the time is the same as it is discussed in V.~for the Tesla coil in more detail, namely it is the difference between two differently damped oscillations with different periods. Like presented there, the coupling $k'=0.6$ $(\tau:\tau'=2)$ of both parts of the receiver thus has to be especially advantageous also here. \textit{Changes of dampings also make no big difference for the maximal amplitude,} as in Fig.~2. \textit{At least much less than for the integral effect},\footnote{The coherer reacts on the maximum amplitude, Rutherford's magnetic indicator probably to the integral effect. Hence, one has to pay more attention on small damping for the latter than for the former.} since that one is proportional to $1:\sigma$, which can be easily obtained from (\ref{eqn:143}). After all, one can see that for the double built (coupled) receiver the inperiodic progressing of the transmitter potential is no further a obstacle. 

 At least, one will obtain the sharpest resonance (although with abandoning the highest possible intensity), if one used a very loose coupled transmitter and a very loose coupled receiver, following M.~Wien, for which the damping

\begin{center}	
\rule[0.3em]{0.3 \columnwidth}{0.5pt}
{\bf Page 560} 
\rule[0.3em]{0.3 \columnwidth}{0.5pt}
\end{center}

\noindent
of the two circuits 1, which contain the high capacity $C_1$, has to be made as small as possible. Furthermore, for sharp resonance it is better to use an indicator reacting on the integral effect than one reacting on the maximum amplitude. Because, in general, at the arrangement the resonance has to be detected the best where the intensity is influenced the strongest from the damping (even in case of resonance).  

It is a question by itself (which will be discussed later), how to most practically scale the antenna in regards to the secondary coil and how to best scale the secondary coil in terms of height, radius and number of windings for the purpose of emission of electromagnetic energy.



\subsection*{Summary of Main Results}\addcontentsline{toc}{subsubsection}{Summary of Main Results}
1. From the combined observation of the maximum-amplitude resonance curve and the integral effects one can find \emph{the damping of both oscillating circuits uniquely, as well as the frequency of the oscillation.}

2. The resonance curve becomes more pronounced, the weaker the coupling between the oscillating circuits is. Further, it is more pronounced for the integral effects than for the maximum amplitude. 

3. The most effective Tesla transformer comes from \emph{one} primary coil and \emph{many} secondary coils, which constitute a body of coils \emph{with a particular} (not yet determined) \emph{ratio of the height to the diameter}. The quantity of secondary coils is limited  through the requirement that the insulation is not punctured and the wire thickness is not too thin. Further, an increase in the number of secondary coils requires a spark-inductor of higher sparking length. Dead (not inductively effective) self-induction of the primary coil is possible to  work around; its wire thickness should be as large as possible and \emph{the coupling between the primary coil and the Tesla coil should be as close as possible to the value $k^2=0.36$} (ratio of the frequencies of 1:2, originating through the coupling). The primary capacitor must bring the primary circuit in resonance with the Tesla 

\begin{center}	
\rule[0.3em]{0.3 \columnwidth}{0.5pt}
{\bf Page 561} 
\rule[0.3em]{0.3 \columnwidth}{0.5pt}
\end{center}

\noindent
coil and should (to protect from electric discharge and residue) be in well-insulated, residue-free dielectric (oil, not glass or air), out of metal contact. 

With weak coupling, it matters only \emph{that the primary capacitance $C_1$ is as large as possible}, irrespective of whether the couplings are achieved through small coils with high winding number $n$, or through large coils with smaller $n$; with strong coupling, high $n$ is somewhat more efficient. --- Within certain limits, the Tesla coil depends little on sparking potential. 

4. With weak coupling, the damping of the primary and secondary circuits has a strong influence on the effectiveness of the Tesla transformers; with stronger coupling (even from $k^2 = 0.16$ on) much less. 

5. With radio telegraphy one finds the sharpest resonance (neglecting the intensity) through weakly coupled and undamped sender and receiver. The latter should include a meter that responds to the integral effect.

6. If one has a weakly coupled (or separated) receiver and a strongly coupled sender, there is no distinct resonance. \emph{First through much stronger coupling in the sender $(k^2 > 0.6)$ can one tune the receiver to the sender. The ratios of the receiver frequency to the frequencies of both }(coordinated)\emph{ oscillating circuits of the sender must then be less than $1:\sqrt{2}$. }

7. If one has two identically built and identically coupled instruments as sender and receiver, then one can achieve a high intensity and moderate precision in the resonance, the latter more from the integral effect than from the maximum amplitude. To achieve the best performance, the coupling in both instruments should be $k^2 = 0.36$. By using a meter sensitive to the maximal amplitude, the outcome depends less on the damping, so that the resonance becomes less precise than if one had measured it with the integral effect.

Giessen, November 1903.
\begin{center}
(Received 25 November, 1903.)
\end{center}


\begin{center}
\rule[0.3em]{0.3 \columnwidth}{1.0pt}
{\bf End of article and translation}
\rule[0.3em]{0.3 \columnwidth}{1.0pt}
\end{center}

\newpage

\section{Discussion}
In the beginning of the article, Drude analyzes a half-wave (or bipolar) Tesla transformer operating near the fundamental (uncoupled) self-resonance frequency of its secondary circuit.  The primary circuit is treated as an ideal lumped-element circuit.  The secondary circuit is a single-layer solenoid without any capacitive loading (no discharge terminals) or ground connection.  

Drude's approach is essentially transmission-line analysis, although he treats inductance using Hopkinson's law, which is described in Sec.~\ref{HLaw}.  For example, (\ref{eqn:8}) and (\ref{eqn:9}) are the Telegrapher's equations for a transmission line with constant, distributed shunt capacitance $\mathfrak{C}_2$ and series resistance $\mathfrak{w}_2$.  He expands $i_2(z)$ and $V_2(z)$ into Fourier series for spatial standing waves along the secondary coil, and retains only the fundamental spatial mode.  In the end, he derives a system of equations which represents an equivalent circuit for a half-wave Tesla transformer.  

In this section, we provide a modern interpretation of this analysis, and extend the results to a quarter-wave Tesla transformer.  We list the errors we noticed in the article, and also describe possible modifications to match the equivalent secondary self-inductance with that of previous work by Drude and others~\cite{drude:1902, hubbard:1917}.  
With these modifications, the predictions of nonreciprocal mutual inductance agree with the results of McGuyer~\cite{mcguyer:2014}. 

\subsection{Technical glossary}
Here is a modern interpretation of some of the language in the translation:  
\begin{description}  \setlength{\itemsep}{1pt} \setlength{\parskip}{0pt} \setlength{\parsep}{0pt}
\item[maximum amplitude, integral effect] (first uses p.~512) two wireless telegraphy detection techniques. 
\item[current strength] (first use p.~513 with $i_1$) current. 
\item[damping coefficient] (first use p.~514 with $w_1$) resistance. 
\item[magnetic field lines] (first use p.~514 with $N_1$) magnetic flux. 
\item[magnetomotive strength] (first use p.~514 with $4\pi i_1$) same as magnetic force (see below). 
\item[magnetic force] (first use p.~515) magnetomotive force (MMF). 
\item[magnetic resistance] (first use pp.~514--5 with $W_{11}$) reluctance. 
\item[law of magnetic circuits] (first use p.~516) Hopkinson's law. 
\item[tube strengths] (first use p.~516 with $N_{12}$) same as magnetic field lines (see above). 
\end{description}

\subsection{Hopkinson's law}\label{HLaw}
Hopkinson's law (or Rowland's law) for magnetic circuits resembles Ohm's law for electric circuits~\cite{wiki:mcircuits}.  In Ohm's law, the electromagnetic force (EMF, or voltage) across some element is equal to the current $I$ passing through it times its electrical resistance $R$: 
\begin{align} 
\text{Ohm's law:   EMF} & = I R.
\end{align}
In Hopkinson's law, the magnetomotive force (MMF) across some element is equal to the magnetic flux $\Phi_\text{m}$ through it times its reluctance $R_\text{m}$:  
\begin{align} 
\text{Hopkinson's law:  MMF} & = \Phi_\text{m} R_\text{m}.
\end{align}
Comparing both laws, we see that the MMF plays the role of an EMF, the magnetic flux $\Phi_\text{m}$ of a current $I$, and the reluctance $R_\text{M}$ of an electrical resistance $R$.

As an example, consider an ideal lumped-element inductor with $N$ turns (not to be confused with Drude's $N$), self-inductance $L$, and current $I$ flowing through it.  The MMF given by Ampere's law is $N I$, and can be thought of as the equivalent current if there was only a single turn ($N=1$).  The magnetic flux $\Phi_\text{m}$ is almost the same as the flux in the definition of self-inductance ($\Phi = L I$), except that for magnetic circuits it is typically the flux per turn, $\Phi_\text{m} = L I / N$. From Hopkinson's law, the reluctance is $R_\text{M} =$ MMF$ / \Phi_\text{m} = N^2 / L$.  For a single-turn inductor ($N=1$), the reluctance is just the reciprocal of the inductance.  

In the translation, Eq.~(\ref{eqn:5}) shows that Drude's MMF has an extra factor of $4 \pi$, likely from using cgs units (with $c=1$).  
Eq.~(\ref{eqn:14}) shows that his MMF is proportional to the number of turns, following Ampere's law.  
Eq.~(\ref{eqn:9}) suggests that his magnetic flux is a flux per turn, following $\Phi_\text{m}$. 
As a result, his reluctance should be of the form $4 \pi R_\text{m}$.

\subsection{Equivalent circuit parameters from the article}
Systems (\ref{eqn:16}) and (\ref{eqn:21}) describes an equivalent circuit for an unloaded, half-wave Tesla transformer.  
With some changes, we can relate this circuit to the conventional equivalent circuit for a Tesla transformer, which is described by the system
\begin{align}
I_\text{p}(t) + C_\text{p} \frac{d V_\text{p}(t)}{dt} &= 0 \label{ECP1} \\
L_\text{p} \frac{d^2 I_\text{p}(t)}{dt^2} + R_\text{p} \frac{d I_\text{p}(t)}{d t} + \frac{1}{C_\text{p}} I_\text{p}(t) + M_\text{ps} \frac{d^2 I_\text{s}(t)}{d t^2} &= 0 \label{ECP2} \\
I_\text{s}(t) + C_\text{s} \frac{d V_\text{s}(t)}{dt} &= 0 \label{ECP3} \\
L_\text{s} \frac{d^2 I_\text{s}(t)}{dt^2} + R_\text{s} \frac{d I_\text{s}(t)}{d t} + \frac{1}{C_\text{s}} I_\text{s}(t) + M_\text{sp} \frac{d^2 I_\text{p}(t)}{d t^2} & = 0.  \label{ECP4}
\end{align}
Here, the first two lines describing the primary circuit follow directly from (\ref{eqn:1}) and (\ref{eqn:16}).  
The primary circuit current $I_\text{p}(t) = i_1$, voltage $V_\text{p}(t) = V_1$, inductance $L_\text{p}=L_{11}$, capacitance $C_\text{p}=C_1$, and resistance $R_\text{p}=w_1$.  

However, the third and fourth lines of (\ref{ECP4}) require some care with factors of 2.  
Following Drude, the secondary voltage $V_\text{s}(t) = V_2$ of (\ref{eqn:21}) is the potential difference between one free end to the middle winding ($V_2^h$), and the secondary current $I_\text{s}(t) = i_2$ of (\ref{eqn:16}) is the middle winding's current ($i_2^0$).  
The third line comes from footnote 13 on p.~519, which gives the capacitance $C_\text{s} = 2 C_2$.  
The fourth line then follows from dividing (\ref{eqn:16}) by two, or (\ref{eqn:21}) by $C_\text{s}$, which gives the inductance $L_\text{s} = L_{22}/2$ and resistance $R_\text{s} = w_2/2$.  

Note, however, that Drude predicts a nonreciprocal mutual inductance, $M_\text{ps} \neq M_\text{sp}$.  From (\ref{eqn:16}) and (\ref{eqn:17}), the effective mutual inductance seen by the primary is $M_\text{ps} = L_{12}$.  Likewise, for the half-wave case treated by Drude, the effective mutual inductance seen by the secondary is $M_\text{sp} = L_{21}/2$.  
Unfortunately, the coefficients $L_\text{s}$ and $M_\text{ps}$ are left in terms of the unknown parameters $\sin (\pi a_1/h)$ and $\sin (\pi a_2/h)$.  

To extend the results to a quarter-wave Tesla transformer, consider removing the $z < 0$ region of the secondary coil.  
Conveniently, the secondary voltage and current definitions still hold and, with the exception of the mutual inductances, most of the circuit parameters are unchanged.  
We will return to this in Sec.~\ref{mod}. 

Finally, note that the above relations follow directly from the translation, and do not account for the errors mentioned in the next section.

\subsection{Errors in the article}\label{errors}
The equations in this translation copy the article.  Errors we noticed are listed here:  
\begin{itemize}  \setlength{\itemsep}{1pt} \setlength{\parskip}{0pt} \setlength{\parsep}{0pt}
\item pp.~516-7:  $N_2$ was used instead of the Fourier term $N_2^0$ in (\ref{eqn:12}) to get (\ref{eqn:15}).   
\item p.~517:  The $W_{11}$ in the expression for $L_{2 2}$ in (\ref{eqn:17}) does not match the $W_{12}$ in (\ref{eqn:15}), and probably should be a $W_{12}$. 
\item p.~518:  After (\ref{eqn:18}), the discussion of $N_{21}$ ignores the contribution from $i_2$.  This seems to be inconsistent with the earlier interpretation of $N_{12}$ and $N_{21}$.  For example, (\ref{eqn:13}) implies $N_{12}=N_{21}$, and (\ref{eqn:6}) gives $N_{12}$ with contributions from both $i_1$ and $i_2$.  
\item p.~518:  The expression for $L_{21}$ in (\ref{eqn:20}) seems to be missing a factor of $1/W_{12}$. 
\end{itemize}

\subsection{Modifications to match $L_\text{s}$ with previous work}\label{mod}
The article gives the equivalent secondary coil self-inductance $L_\text{s} = L_{22}/2$ in terms of the unknown parameters $\sin (\pi a_1/h)$ and $\sin (\pi a_2/h)$ in (\ref{eqn:17}).  
This result differs from previous work by Drude (and others) for quarter-wave Tesla transformers, which give the result 
\begin{align}\label{previousLs}
L_\text{s} =  \left( \frac{2}{\pi} \right) l H, 
\end{align} 
where $H=h/2$ is the height of the secondary coil and $l$ is a distributed series inductance, such that $l H = L_\text{s}^\text{(dc)}$ is the low-frequency self-inductance of the coil \cite{drude:1902, hubbard:1917}.  
(However, modern equivalent circuits for solenoids typically use the this low-frequency self inductance instead of (\ref{previousLs})~\cite{hubbard:1917}.) 
Interestingly, the two parameters in (\ref{eqn:17}) that are expressed in terms of distributed quantities share a similar form,  
\begin{align}\label{curious}
C_\text{s} =  2 C_2 = \left( \frac{2}{\pi} \right) \mathfrak{C}_2 H \quad \text{ and } \quad R_\text{s} =  \frac{w_2}{2} = \left( \frac{2}{\pi} \right)  \mathfrak{w}_2 H.  
\end{align}
The similarity of (\ref{previousLs}) with (\ref{curious}) is suspicious, and raises the question of why Drude did not obtain the same result (\ref{previousLs}) again for $L_\text{s}$.  
One possible explanation is that Drude's use of magnetic fluxes $N$ with Fourier series is incorrect. (It is at least inconsistent.)  
What follows is a modified derivation which recovers the previous result (\ref{previousLs}).  

First, note that there was an error going from (\ref{eqn:12}) to (\ref{eqn:15}):  $N_2$ was used for (\ref{eqn:15}) in place of the Fourier term $N_2^0$ in (\ref{eqn:12}).  
To fix this, we need to calculate the $N_{2}^0$ of (\ref{eqn:11}).  We can suppose that $N^0_{2} = N^0_{22} + N^0_{21}$, 
\begin{align} 
N_{22}^0 &= \frac{4\pi}{W_{22}} \left( \frac{2}{h} \right) \int_{-h/2}^{h/2} n \, i_2(z) \cos \left( \frac{\pi z}{h} \right)dz = \frac{4\pi n i_2^0}{W_{22}}, \label{N220} \\
\text{and that}  \quad
N_{21}^0 &= \frac{4\pi}{W_{21}} \left( \frac{2}{h} \right) \int_{-h/2}^{h/2} [i_1 + n\, i_2 (z) ] \cos \left( \frac{\pi z}{h} \right)dz = \frac{16 i_1 + 4 \pi n i_2^0}{W_{21}}.  \label{N221}
\end{align}
(Note that the $N_{21}^0$ on p.~518 is a different quantity.)  As a result, in (\ref{eqn:15}) we should have the substitutions
\begin{align} \label{subs} 
8 n^2 \left( \frac{\sin (\pi a_1/h)}{W_{21}} + \frac{\sin (\pi a_2/h)}{W_{22}} \right) \longrightarrow 4 \pi n^2 \left( \frac{1}{W_{21}} + \frac{1}{W_{22}} \right) \quad \text{ and } \quad \frac{4 \pi n}{W_{12}} \longrightarrow \frac{16 n}{W_{21}}.  
\end{align}
The first substitution changes the $L_{22}$ in (\ref{eqn:17}) to  
\begin{align} \label{L22fix} 
L_{22} =  \left( \frac{2}{\pi} \right) 4 \pi n^2 \left( \frac{1}{W_{21}} + \frac{1}{W_{22}} \right).  
\end{align}

However, to treat a quarter-wave Tesla transformer, (\ref{N220}) and (\ref{N221}) should be changed as follows:  the total number of turns $n$ should be reduced to $n'=n/2$, the integrals should be restricted to $[0, H]$, where $H = h/2$, and the pre-factors $2/h$ replaced with $2/H$.  With these changes, (\ref{N220}) and (\ref{N221}) become 
\begin{align} 
N_{22}^0 & = \frac{4\pi}{W_{22}} \left( \frac{2}{H} \right) \int_{0}^{H} n' i_2(z)  \cos \left( \frac{\pi z}{h} \right)dz = \frac{4\pi n' i_2^0}{W_{22}} \label{N220p} \\
\text{and} \quad 
N_{21}^0 & = \frac{4\pi}{W_{21}} \left( \frac{2}{H} \right) \int_{0}^{H} [ i_1 + n' i_2(z) ] \cos \left( \frac{\pi z}{h} \right)dz = \frac{16 i_1 + 4 \pi n' i_2^0}{W_{21}}.  \label{N221p} 
\end{align}
Comparing (\ref{N220}-\ref{N221}) with (\ref{N220p}-\ref{N221p}), we find from (\ref{L22fix}) that the $L_\text{s}$ for a quarter-wave Tesla transformer is  
\begin{align} 
L_\text{s} = \frac{L_{22}}{2} =   \left( \frac{2}{\pi} \right) 4 \pi (n')^2 \left( \frac{1}{W_{21}} + \frac{1}{W_{22}} \right).	\label{LsFinally}
\end{align}
Finally, inspecting the form of $L_\text{p} = L_{11}$ in (\ref{eqn:17}), we can suppose that $l H = L_\text{s}^\text{(dc)} = 4 \pi ( 1/W_{21} + 1/W_{22})$, so we see that the result (\ref{LsFinally}) is of the desired form (\ref{previousLs}). 

These modifications also change the results for mutual inductance.  
For the half-wave case, the second substitution in (\ref{subs}) changes the $L_{21}$ in (\ref{eqn:17}) to  
\begin{align} 
L_{21} = \left( \frac{2}{\pi} \right) \frac{16 n}{W_{21}}.  
\end{align}
Note that this also holds for the quarter-wave case, since the factor of $n$ here comes from the turn density $n/h = N/H$ introduced in (\ref{eqn:12}).  
Therefore, the mutual inductance for both cases is 
\begin{align}	\label{Mspfinal}
M_\text{sp} = \frac{L_{21}}{2} = \left( \frac{2}{\pi} \right) \frac{8 n}{W_{21}} = \left( \frac{4}{\pi} \right) \frac{8 n'}{W_{21}}.
\end{align}
Focusing now on the primary, in place of the derivation of $N_{12}$ on p.~515, we can suppose that 
\begin{align} 
N_{12} = \frac{4\pi}{W_{12}} \left[ i_1 +  \frac{n}{h} \int_{-h/2}^{h/2}  i_2(z) dz \right] = \frac{4 \pi i_1 + 8 n i_2^0}{W_{12}} 
\end{align}
for the half-wave case. As a result, in (\ref{eqn:7}) we should have the substitution 
\begin{align}
\frac{8n}{W_{12}} \sin \left(\frac{\pi a_1}{h} \right) \longrightarrow \frac{8n}{W_{12}}, 
\end{align}
which changes the $L_{12}$ in (\ref{eqn:17}) to 
\begin{align} \label{L12fix} 
L_{12} = \frac{8 n}{W_{12}}. 
\end{align}
However, following the changes described above, for the quarter-wave case we should have  
\begin{align} 
N_{12} = \frac{4\pi}{W_{12}} \left[ i_1 +  \frac{n'}{H} \int_{0}^{H}  i_2(z) dz \right] = \frac{4 \pi i_1 + 8 n' i_2^0}{W_{12}},   
\end{align}
which again leads to the same result (\ref{L12fix}) if $n$ is replaced with $n'$. 
Therefore, the mutual inductance 
\begin{align}	\label{Mpsfinal}
M_\text{ps} = L_{12} = \frac{8 n}{W_{21}}  \, \text{ for the half-wave case, or } \, 
\frac{8 n'}{W_{21}}  \, \text{ for the quarter-wave case}.
\end{align}

Using (\ref{Mpsfinal}) with (\ref{Mspfinal}), we find that the ratio of the mutual inductances 
\begin{align}
\frac{ M_\text{ps} }{ M_\text{sp} } &= \frac{2L_{12}}{L_{21}} = \frac{\pi}{2}  \, \text{ for the half-wave case, and } \, 
\frac{\pi}{4}  \, \text{ for the quarter-wave case}.
\end{align}
We see that Drude's prediction on p.~518 that $L_{12} < L_{21}$ still holds, though the unknown parameter $\sin(\pi a_1/h)$ has been eliminated.  
These ratios of mutual inductances agree with the results of McGuyer~\cite{mcguyer:2014}.

\clearpage


\end{document}